%% file: main.tex
\PassOptionsToPackage{dvipsnames,table,xcdraw}{xcolor}
\documentclass[sigconf]{acmart}
\AtBeginDocument{%
  }
\setcopyright{acmlicensed}
\copyrightyear{2025}
\acmYear{2025}
\acmConference[CCS '25]{Proceedings of the 2025 ACM SIGSAC Conference on Computer and Communications Security}{October 13--17, 2025}{Taipei, Taiwan}
\acmBooktitle{Proceedings of the 2025 ACM SIGSAC Conference on Computer and Communications Security (CCS '25), October 13--17, 2025, Taipei, Taiwan}
\acmPrice{15.00}
\acmISBN{978-1-4503-XXXX-X/18/06}

\usepackage{url}                
\hypersetup{                    
  colorlinks,
  linkcolor={blue!70!black},
  citecolor={red!70!black},
  urlcolor=RoyalPurple
}
\usepackage{hyperref}         

\usepackage{hhline}

\usepackage{soul}

\newcommand{\revisioncolor}{black}
\newcommand{\tcolor}[1]{\textcolor{\revisioncolor}{#1}}

\newcommand{\revOne}[1]{\textcolor{\revisioncolor}{#1}}
\newcommand{\rcolor}[1]{\textcolor{\revisioncolor}{#1}}

\newcommand{\revTwo}[1]{\textcolor{\revisioncolor}{#1}}

\usepackage{tikz}

\newcommand{\circlednum}[1]{
\tikz[baseline={([yshift=-3.3pt]current bounding box.center)}]{ 
\node[shape=circle,draw,inner sep=1pt, fill=black, text=white] (char) {\scriptsize #1};
}}
\newcommand{\circlednuminline}[1]{
\hspace{-3pt}\tikz[baseline={([yshift=-3.3pt]current bounding box.center)}]{
\node[shape=circle,draw,inner sep=1pt, fill=\revisioncolor, text=white, anchor=base west] (char) {\scriptsize #1};
}}
\usepackage{makecell}
\usepackage{tabularx}
\usepackage{tabularray}
\usepackage{threeparttable}

\newcommand{\customsize}{\fontsize{8.5}{10.5}\selectfont}
\newenvironment{smalltabularx}{\customsize \tabularx}{\endtabularx}
\newcommand{\customtinysize}{\fontsize{8}{10}\selectfont}
\newenvironment{tinytabularx}{\customtinysize \tabularx}{\endtabularx}

\newcolumntype{C}{>{\centering\arraybackslash\hsize=.5\hsize\linewidth=\hsize}X}

\usepackage{siunitx} 
\usepackage{fp}
\newcommand{\perc}[1]{%
\footnotesize
  \FPifneg{#1}
    \textcolor{blue!70!black}{\footnotesize $\downarrow$\num[round-mode=places,round-precision=1]{\stripminus #1}\%}
  \else
    \textcolor{blue!70!black}{\footnotesize \textbf{$\uparrow$\num[round-mode=places,round-precision=1]{#1}\%}}
  \fi
}

\newcommand{\percplus}[1]{%
  \FPifneg{#1}%
    \textcolor{blue!70!black}{\footnotesize$\downarrow$\num[round-mode=places,round-precision=2]{\stripminus #1}\%}%
  \else%
    \textcolor{blue!70!black}{\footnotesize\textbf{$\uparrow$\num[round-mode=places,round-precision=2]{#1}\%}}%
  \fi%
}

\newcommand{\stripminus}[1]{\expandafter\the\numexpr-#1}

\usepackage{booktabs}
\usepackage{subfig}
\usepackage{graphicx}

\usepackage{amsmath}
\usepackage{amssymb}

\usepackage{wasysym}
\DeclareMathOperator*{\argmax}{argmax}

\usepackage[framemethod=TikZ]{mdframed}
\mdfdefinestyle{mystyle}{
  linecolor=black,
  outerlinewidth=.1pt,
  skipabove=-18pt,
  skipbelow=1pt,
  innertopmargin=-2pt,
  innerbottommargin=0.2pt,
  innerrightmargin=5pt,
  innerleftmargin=5pt,
  rightmargin=5pt,
  leftmargin=5pt,
  font=\customtinysize,
}

\mdfsetup{style=mystyle}

\usepackage{multirow}

\usepackage{enumitem}
\setlist[itemize]{leftmargin=*,itemsep=-0.1em,topsep=0.2em, partopsep=0.2em}

\usepackage[nameinlink,noabbrev]{cleveref}

\usepackage{etoolbox}
\makeatletter
\patchcmd{\hyper@makecurrent}{%
    \ifx\Hy@param\Hy@chapterstring
        \let\Hy@param\Hy@chapapp
    \fi
}{%
    \iftoggle{inappendix}{
        \@checkappendixparam{chapter}%
        \@checkappendixparam{section}%
        \@checkappendixparam{subsection}%
        \@checkappendixparam{subsubsection}%
        \@checkappendixparam{paragraph}%
        \@checkappendixparam{subparagraph}%
    }{}%
}{}{\errmessage{failed to patch}}

\newcommand*{\@checkappendixparam}[1]{%
    \def\@checkappendixparamtmp{#1}%
    \ifx\Hy@param\@checkappendixparamtmp
        \let\Hy@param\Hy@appendixstring
    \fi
}
\makeatletter

\newtoggle{inappendix}
\togglefalse{inappendix}

\apptocmd{\appendix}{\toggletrue{inappendix}}{}{\errmessage{failed to patch}}

\usepackage[normalem]{ulem}

\newcommand{\framework}{\textsc{Dream}}
\newcommand{\bfnoindent}[1]{\noindent\textbf{#1.}} 
\usepackage{balance}

\renewcommand{\footnotesize}{\scriptsize}
\renewcommand{\footnoterule}{%
  \kern -2pt                              
  \hrule width 1in height 0.3pt           
  \kern 2pt                             
}

\begin{document}

\setlength{\aboverulesep}{-.5pt}
\setlength{\belowrulesep}{-.5pt}

\title{Combating Concept Drift with Explanatory Detection and Adaptation for Android Malware Classification} 

\author{Yiling He}
\authornote{This work was conducted during the author's PhD studies at Zhejiang University.}
\affiliation{%
  \institution{University College London}
  \city{London}
  \country{United Kingdom}
  }
\email{yiling-he@ucl.ac.uk}

\author{Junchi Lei}
\affiliation{%
  \institution{Zhejiang University}
  \city{Hangzhou}
  \country{China}
}
\email{junchilei@zju.edu.cn}

\author{Zhan Qin}
\affiliation{%
  \institution{Zhejiang University}
  \city{Hangzhou}
  \country{China}
}
\email{qinzhan@zju.edu.cn}

\author{Kui Ren}
\affiliation{%
  \institution{Zhejiang University}
  \city{Hangzhou}
  \country{China}
}
\email{kuiren@zju.edu.cn}

\author{Chun Chen}
\affiliation{%
  \institution{Zhejiang University}
  \city{Hangzhou}
  \country{China}
}
\email{chenc@zju.edu.cn}

\renewcommand{\shortauthors}{Yiling He et al.}


\begin{abstract}
\input{abstract}
\end{abstract}



\maketitle


%

\section{Introduction}
\input{intro}

\input{background}

\input{revision/overview_new}

\input{method}

\section{System Evaluation}

\input{evaluation}

\section{Extended Evaluation} \label{eva:extend}
\input{empirical}

\input{discussion}

\input{related}

\section{Conclusion}
\input{conclusion}






\bibliographystyle{ACM-Reference-Format}
\bibliography{reference/dream, reference/finer}

\appendix

\input{appendix}
\end{document}

%% file: abstract.tex
Machine learning-based Android malware classifiers achieve high accuracy in stationary environments but struggle with concept drift.
The rapid evolution of malware, especially with new families, can depress classification accuracy to near-random levels. 
Previous research has largely centered on detecting drift samples, with expert-led label revisions on these samples to guide model retraining.
However, these methods often lack a comprehensive understanding of malware concepts and provide limited guidance for effective drift adaptation, leading to unstable detection performance and high human labeling costs.

To combat concept drift, we propose \framework{}, a novel system that improves drift detection and establishes an explanatory adaptation process.
\tcolor{Our core idea is to integrate classifier and expert knowledge within a unified model.
To achieve this, we embed malware explanations~(or concepts) within the latent space of a contrastive autoencoder, while constraining sample reconstruction based on classifier predictions. 
This approach enhances classifier retraining in two key ways: 1)~capturing the target classifier’s characteristics to select more effective samples in drift detection and 2)~enabling concept revisions that extend the classifier’s semantics to provide stronger guidance for adaptation.
Additionally, \framework{} eliminates reliance on training data during real-time drift detection and provides a behavior-based drift explainer to support concept revision.}
Our evaluation shows that \framework{} effectively improves the drift detection accuracy and reduces the expert analysis effort in adaptation across different malware datasets and classifiers.
\tcolor{Notably, when updating a widely-used \textsc{Drebin} classifier, \framework{} achieves the same accuracy with $76.6\%$ fewer newly labeled samples compared to the best existing methods.}

%% file: intro.tex
Android malware classification is continually challenged by concept drift~\cite{pendlebury2019tesseract}.
As cyber attackers constantly devise evasion techniques and varied intents~\cite{avllazagaj2021malware}, the evolving nature of malware behaviors can rapidly alter the patterns which classifiers rely on.  
Consequently, static machine learning models trained on historical data face a significant drop in performance and become incapable of handling unseen families~\cite{he2023msdroid}.

To combat malware concept drift, current state-of-the-art solutions leverage active learning~\cite{hsu2015active}, comprising two primary stages. 
In the \textit{drift detection} stage, new test samples that exhibit signs of drift are periodically selected. Most existing research has focused on improving this stage using techniques like statistical analysis~\cite{jordaney2017transcend, barbero2022transcending} or contrastive learning~\cite{yang2021cade, chen2023continuous} to identify atypical data points. The subsequent \textit{drift adaptation} stage follows a standard approach: the identified drifting samples are labeled by malware analysts and then added to the training set for classifier retraining~\cite{zhang2020enhancing}.


Existing methods for drift detection fall short in two main aspects.
Firstly, they often falsely neglect the patterns which the targeted classifier depends on.
For example, the CADE detector~\cite{yang2021cade} leverages an independent autoencoder to learn a distance function, identifying drifts by the distances with training data.
Such misalignment can lead to inefficiency in detecting model-specific drifts, especially when dealing with complex classifier feature spaces, as validated by our experiments in~\autoref{sec:eva_detection}. 
Secondly, existing methods often rely on constant access to training data during the testing phase, such as for querying reference uncertainties~\cite{barbero2022transcending, jordaney2017transcend} or searching for nearest neighbors~\cite{chen2023continuous}. 
Such practices bring practical issues in local deployment scenarios, where managing large training datasets leads to storage and security concerns~\cite{ren2023demistify}.

In the drift adaptation stage, the common label-centric retraining strategy poses significant challenges to human efforts. 
A clear discrepancy exists between the extensive analysis required for assigning labels and the limited information actually utilized by the model. 
While analysts use static/dynamic program analysis to extract rich reasoning insights~\cite{antovaniAFB22, humanvsmachine}, these insights are largely disregarded as the model only considers the revised labels.
This mismatch limits the effectiveness of model updates, with mislabeling further complicating improvements~\cite{wu2023grim, pirch2021tagvet}.
Consequently, analysts often need to label a large number of samples to ensure accurate updates.


To address these challenges, our high-level idea is to design a system that aligns closely with the classifier knowledge while supporting human revision of malware behavioral concepts.
This alignment facilitates the two integral processes:  
1)~in drift detection, deviations are identified by assessing the reliability of malware concepts under the supervision of the classifier; 
2)~in drift adaptation, experts are provided with interfaces to directly revise the deviated concepts, enabling these adjustments to inversely update the classifier.

We develop the system, called \framework{}, distinguished by model-sensitive \uline{dr}ift detection and \uline{e}xplanatory \uline{a}daption for Android \uline{m}alware classification.
Specifically, we train an autoencoder with model sensitive concept learning.
To embed malware concepts, we combine supervised concept learning~\cite{jia2013visual} with unsupervised contrastive learning~\cite{chen2020contrastive}: elements in the latent space are representative of behavioral concepts that can be labeled, such as remote control and stealthy download.
For drift detection, \framework{} integrates classifier knowledge and achieves data-independent testing through a unique concept reliability measure, grounded in the intuition that reconstructed samples from reliable concepts should yield consistent model predictions with the original samples.

For adaptation, we envision that human intelligence can not only be utilized for feedback on prediction labels but also for the explanations~\cite{teso2019explanatory}.
Rather than creating an external explanation module for the classifier, we utilize concepts embedded within our detector.
This integration offers dual benefits: high-level explanation abstraction for human understanding~\cite{he2023finer} and immediate revision impact on evolving concepts.
Furthermore, \framework{} effectively harnesses human intelligence with a joint update scheme. 
The classifier's retraining process is assisted by the detector, allowing for the utilization of both malware labels and behavioral explanations.


We evaluate \framework{} with $2$ distinct Android malware datasets \cite{yang2021cade, wang2022malradar} and $3$ state-of-the-art classifiers~\cite{arp2014drebin, mariconti2016mamadroid, mclaughlin2017deep}. 
Results demonstrate that \framework{} outperforms $3$ existing drift detection methods~\cite{barbero2022transcending, yang2021cade, chen2023continuous} in malware classification, achieving an average improvement of \text{11.5\%, 12.0\%, and 13.6\%} in terms of the AUC~\footnote{\hypersetup{citecolor=black}The AUC specifically refers to the Area Under the Receiver Operating Characteristic (ROC) Curve~\cite{fan2006roc} in this paper.} metric.
For drift adaptation, \framework{} notably enhances the label-centric retraining approach. 
It achieves improved F1-scores of \text{168.4\%, 73.5\%, 22.2\%, 19.7\%, and 6.0\%} \revOne{over the benchmark under the same data constraints} when the human analysis budget is set at $10$, $20$, $30$, $40$, and $100$, respectively.
A standout observation is the increased advantage at lower human labeling costs. 
For instance, analysts can analyze over \text{80\%} fewer samples while still achieving a test accuracy of $0.9$ for the updated classifier.

\begin{table}[t]
\centering
\begin{threeparttable}
    \begin{smalltabularx}{\linewidth}{cCCCCCCC}
        \toprule
        & \multicolumn{2}{c}{\textbf{Detection}} & \multicolumn{2}{c}{\textbf{Adaptation}} & \multicolumn{3}{c}{\textbf{Application}} \\ 
        \cmidrule(r){2-3} \cmidrule(lr){4-5} \cmidrule(l){6-8} 
        & \textbf{MS} & \textbf{DA} & \textbf{Act.} & \textbf{Exp.} & \textbf{Inter} & \textbf{Intra} & \textbf{Agn.} \\ 
        \midrule
        Trans.~\cite{barbero2022transcending} & \LEFTcircle & \Circle & \Circle & \Circle & \LEFTcircle & \LEFTcircle & \CIRCLE \\
        CADE~\cite{yang2021cade} & \Circle & \LEFTcircle & \CIRCLE & \LEFTcircle & \LEFTcircle & \CIRCLE & \CIRCLE \\
        HCC~\cite{chen2023continuous} & \CIRCLE & \Circle & \CIRCLE & \Circle & \Circle & \CIRCLE & \Circle \\ \midrule
        \framework{} & \CIRCLE & \CIRCLE & \CIRCLE & \CIRCLE & \CIRCLE & \CIRCLE & \CIRCLE \\
        \bottomrule
    \end{smalltabularx}
    \begin{tablenotes}
    \scriptsize
        \item[a] \CIRCLE=true, \Circle=false, \LEFTcircle=partially true.
        \item[b] Detection methods can be distinguished by model sensitivity~(MS) and data autonomy~(DA). Adaptation techniques can be proactive and explanatory. Different support in the detection and proactive adaptation phases includes inter-class drifts, intra-class drifts, and model-agnostic applications.
    \end{tablenotes}
    \caption{Comparison to existing work in terms of the detection method, adaptation method, and the applicable scope.}
    \label{tab:related}
\end{threeparttable}
\end{table}

\bfnoindent{Contributions} 
This paper has three main contributions.
\begin{itemize}
    \item We revisit existing drift detection methods in malware classification, identifying necessity for model sensitivity and data autonomy. We design a novel drift detector to meet the two needs, enriched with the integration of malware behavioral concepts.
    \item We initiate an explanatory concept adaptation process, enabling expert intelligence to revise malware labels and concept explanations. Through a joint update of both the classifier and our drift detector, we harness the synergy between human expertise and automated detection.
    \item We implement \framework{} and evaluate it across different datasets and classifiers. Experimental results show that \framework{} significantly outperforms existing drift detectors and the conventional retraining-based adaptation approach.
\end{itemize}

%% file: background.tex
\begin{figure}[tb]
    \centering
    \includegraphics[width=.96\linewidth]{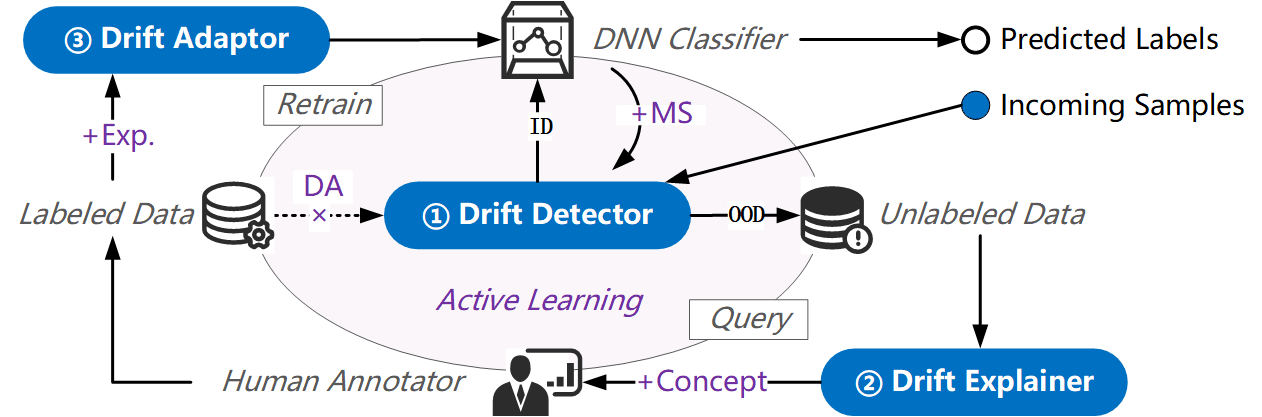}
    \caption{Active learning framework for concept drift mitigation: standard components and operations in common active learning~(grey, italic), specialized components for concept drift~(blue boxes), and new features of our method compared to existing works~(purple).}
    \label{fig:active-learning}
\end{figure}

\section{Background} \label{sec:background}


\input{revision/background_new}

%% file: revision/background_new.tex
In this section, we delve into the background of malware concept drift. 
\tcolor{We begin by discussing the two primary types of drift in learning-based malware classifiers, particularly in the context of Android. 
Next, we introduce existing work for mitigating the drift, highlighting key studies for dynamic adaptation in this domain.} 

\subsection{\tcolor{Android Malware Concept Drift}}

\tcolor{The widespread adoption of Android, which holds a dominant mobile market share of $71.31\%$ as of April 2024~\cite{iqubal2024android}, has made it an attractive target for cybercriminals~\cite{chen2023deuedroid}. 
With over $3$ million new malware samples detected every month~\cite{AV-ATLAS}, the platform faces a rapidly evolving threat landscape. This dynamic environment, driven by attackers seeking to evade~\cite{wang2021beyond, jiang2020aomdroid}, poses significant challenges for both deep \revOne{learning-based} malware detection and classification systems~\cite{li2021robust, gong2020experiences, fan2016frequent, rabadi2020advanced}. 
These systems are particularly vulnerable to \textit{concept drift}, a phenomenon where the statistical properties of data change over time, leading to a degradation in model performance~\cite{guerra2022concept, pendlebury2019tesseract}.}

Concept drift in malware manifests in two primary forms, \tcolor{with the ever-evolving nature of Android malware providing prominent examples}:
1)~\textit{Intra-class drift} refers to changes within existing malware families, such as the emergence of new variants. \tcolor{For example, the information-stealing Android malware \texttt{Xavier} produced $5$ different versions within eight months of its discovery~\cite{xu17report}. }
This type of drift primarily impacts the binary detection tasks, \tcolor{with studies showing that models for Android malware can degrade to near-random levels after two years without updates~\cite{meng2024detecting, he2023msdroid};}
2)~\textit{Inter-class drift}, \tcolor{which is the focus of this paper}, involves the emergence of entirely new malware families. \tcolor{The rapid emergence is exemplified by the identification of $10$ new families of Android banking malware alone in 2023, each targeting critical financial applications globally with different capabilities~\cite{bleepingcomputer2023android}. }
\tcolor{This type of drift is pressing, as it requires timely updates to the multi-class classification models to ensure a more granular response to malware~\cite{gandotra2014malware}.}



\begin{figure*}[tbph]
    \centering
    \subfloat[Latent Space of Detector]{%
    \includegraphics[width=0.49\linewidth]{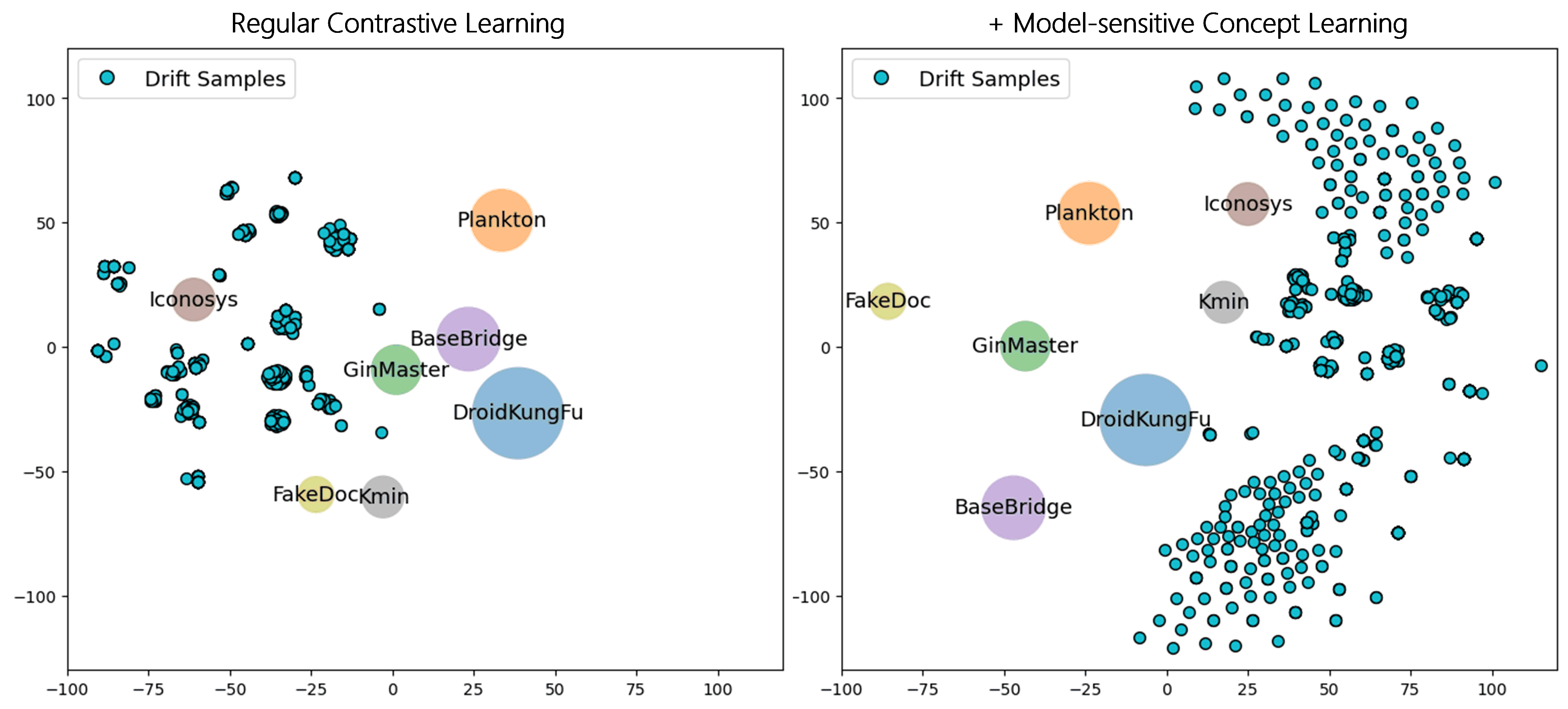}%
    \label{fig:latent_space_detector}}%
    \hfill
    \subfloat[Activation Embedding of Classifier]{%
    \includegraphics[width=0.49\linewidth]{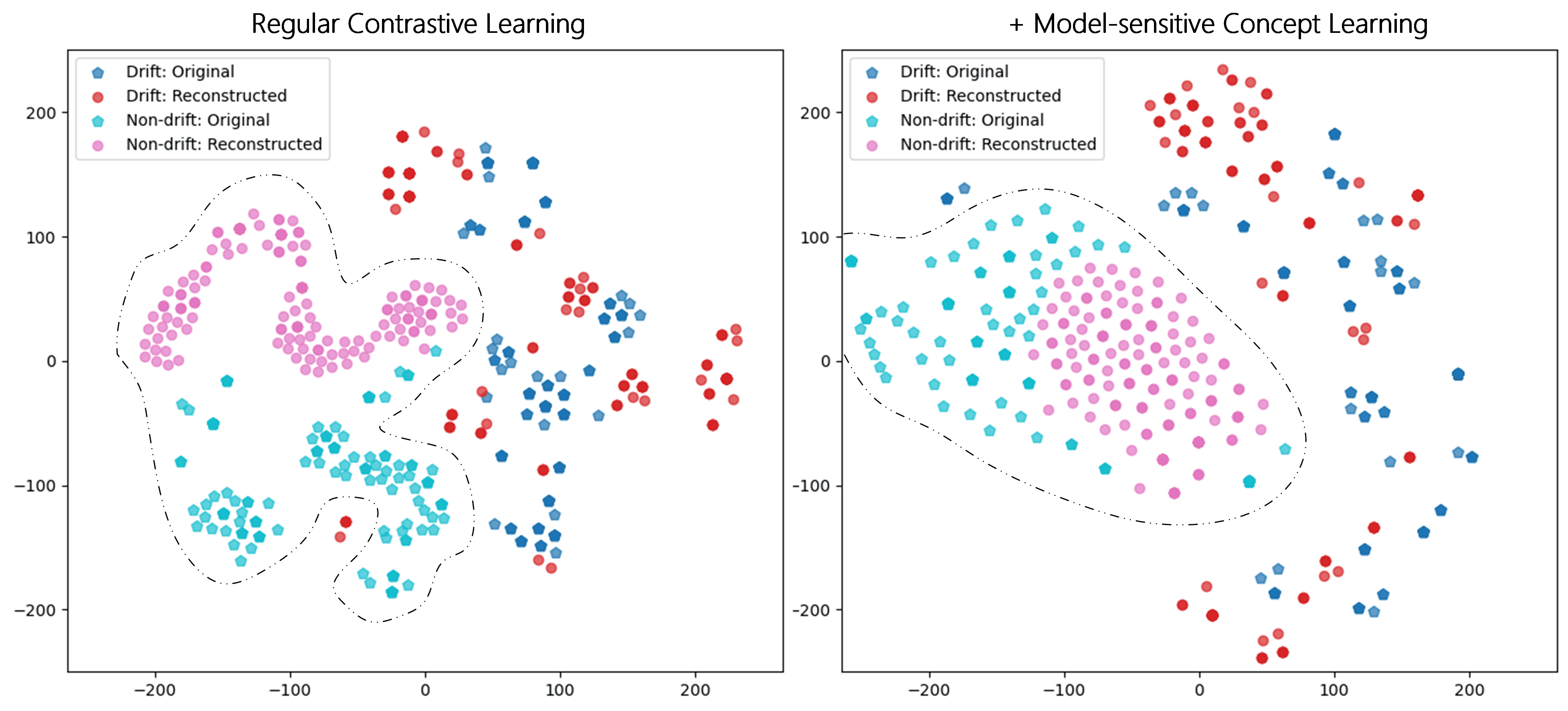}%
    \label{fig:activation_embedding_classifier}}%
    \caption{\tcolor{Model sensitive concept learning: how concepts are learned by the detector and used by the classifier.}}
    \label{fig:embedding_insights}
    \vspace{-.8em}
\end{figure*}

\subsection{\tcolor{Mitigating Malware Concept Drift}}

\tcolor{Traditional approaches to mitigating malware concept drift often lack support for online scenarios that require continuous adaptation.} 
\tcolor{These methods typically involve redesigning feature spaces, followed by retraining from scratch using the new features~\cite{barddal2017survey, ceschin2023fast}. The key challenge in this process is identifying a feature set that remains relevant as the malware evolves. A recent work proposes drift forensics~\cite{chow2023drift}, adopting multiple retraining and feature attribution calculations across different dataset divisions. 
While their in-depth analysis provides valuable insights for feature selection in a binary feature space~\cite{arp2014drebin}, their method relies on ground truth labels and become infeasible for complex classifiers.
Designing robust feature spaces remains an open research question.
}
\tcolor{To address online scenarios, recent research has adopted a two-stage paradigm: determining when to take action (\textit{drift detection}) and what action to take (\textit{drift adaptation}).}

\bfnoindent{Drift Detection} 
The detection stage involves identifying drift by selecting ambiguous samples. Much of the existing research focuses on this stage, drawing inspiration from Out-of-Distribution (OOD) detection methods in the machine learning domain~\cite{yang2021generalized}. For example, \textit{uncertainty estimation} gauges a model's prediction confidence, with a naive approach for DNN being the probability. 
However, probability alone can be misleading, as overfitted models may assign high softmax values to unfamiliar data, falsely indicating high confidence~\cite{hendrycks2016baseline, pearce2021understanding}. As a result, many researchers have shifted towards \textit{nonconformity scoring}, which evaluates how unusual new data is compared to a calibration set~\cite{jordaney2017transcend, barbero2022transcending}. 
Notably, \textit{contrastive learning} has shown promise in modeling an effective distance function for this purpose~\cite{yang2021cade, chen2023continuous}.

\bfnoindent{Drift Adaptation} 
Once drift is detected, the goal is to properly handle the identified samples. A conservative approach involves withholding ambiguous samples for expert analysis to minimize misclassification risks~\cite{jordaney2017transcend, barbero2022transcending}. 
While this strategy provides a safeguard against immediate threats, it lacks long-term resilience. 
Alternatively, \textit{active learning}~(AL) inspired strategies introduce a labeling budget to prioritize a subset of drifting samples for humans to label.
These labeled samples are then integrated into the retraining process, enabling the model to adapt dynamically to new data distributions~\cite{chen2023continuous}.
To streamline human labeling in active learning, there’s growing interest in \textit{drift explanation}~\cite{yang2021cade, hananomaly23owad}, but how these explanations can help update the classifiers remains underexplored.


\bfnoindent{Summary of Key Studies}
Advancements in drift detection and adaptation techniques are essential for combating malware drifts dynamically.  
As summarized in \autoref{tab:related}, three notable existing studies have contributed significantly to the fields~\tcolor{(see more discussions in~\autoref{sec:app_existing}).
Among these, Transcendent focuses on drift detection by refining the calibration process and employs a conservative adaptation approach, whereas CADE and HCC use contrastive learning for drift detection and active learning for adaptation.
Notably, HCC specializes in intra-class drift by incorporating a hierarchical structure into the contrastive loss, and its drift detector has an inherent design that is incompatible with off-the-shelf classifiers.
This paper targets a similar scope of applicability as CADE but also addresses inter-class adaptation, a challenging aspect not considered in these existing works.}

%% file: revision/overview_new.tex
\section{Motivation and Research Scope} \label{sec:motivation}

\tcolor{In this section, we identify the limitations of current drift detection methods and active learning-based drift adaptation strategies, followed by a discussion of our research scope.}

\begin{figure*}[tbhp]
    \centering
    \includegraphics[width=\linewidth]{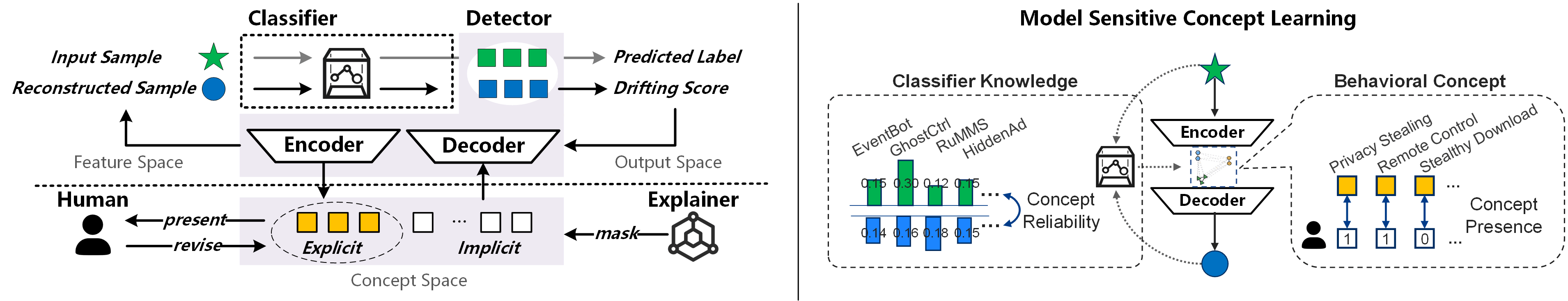} 
    \caption{\tcolor{Overview~(left) and design insights~(right) of our \framework{} system. To enhance contrastive autoencoder-based drift detector, we employ model sensitive concept learning: \framework{} enriches the latent space with concepts of malicious behaviors, and it learns those concepts along with classifier knowledge.}}
    \label{fig:design_insight}
\end{figure*}





\subsection{Limitations on Current Detection Methods}

Current inter-class drift detectors often lack \textit{model sensitivity} and \textit{data autonomy}~(detailed in~\autoref{sec:formalized_detectors}), resulting in ineffective drift detection. 

\textbf{(L1)}
Firstly, these detectors are typically trained independently of the classifier, which should have been the subject of drift. 
For instance, the contrastive learning-based methods~\cite{yang2021cade, chen2023continuous} use an unsupervised approach to learn a distance metric with training data, without fully considering the specific characteristics of the classifier. 
This unguided training leads to a \textit{poorly modeled outlier space}: as illustrated in the left side of Figure~\ref{fig:latent_space_detector}, drift samples (representing a new malware family \texttt{FakeInstaller}) often become entangled with existing classes in the latent space, making it difficult to distinguish between them effectively.

\textbf{(L2)}
Secondly, current detectors rely on training data to compute drift scores in operational phase.
This dependence could be \textit{impractical and inefficient}—it requires constant access to potentially large datasets, posing challenges in terms of security and scalability~\cite{ren2023demistify}. 
Moreover, we find that over-reliance on static training data risks overfitting, causing unseen samples from existing families to be misclassified as drift, further diminishing detection effectiveness. 

\subsection{Challenges in AL-based Adaptation} 


Active learning is employed to dynamically update classifier predictions in drift adaptation, involving human analysts in the process. 
To ensure the updated model is accurate, analysts assign new labels to a contextually determined number of drift samples~(\textit{labeling budget}), which are then incorporated into the original training data for retraining. 
This adaptation stage presents two challenges regarding the effectiveness and efficiency of human efforts.

\textbf{(C1)}
One major challenge is minimizing the labeling budget needed to achieve target accuracy in the updated model. 
The common retraining approach is inherently inefficient, as it relies solely on the final family label provided by analysts, overlooking the valuable reasoning insights they apply during analysis. 
For example, analysts often derive their understanding about malware family from observing specific malicious behaviors through static and dynamic analysis~\cite{humanvsmachine}, which offer valuable context beyond a simple label.
The \textit{failure to incorporate rich behavioral explanations} limits potential improvements in model updating. 
Consequently, this inefficiency increases the labeling burden, requiring more samples to be precisely analyzed~\cite{wu2023grim, pirch2021tagvet}.

\textbf{(C2)}
During label assignment, reducing expert effort in malware analysis remains a challenge~\cite{burk22Decomperson, antovaniAFB22}.
Current drift explanations fall short as generated at the feature level, where intelligibility issues exist especially for malware classifiers~\cite{he2023finer}.
More importantly, unlike explanations for non-drift data~\cite{guo2018lemna}, drift explanations emphasize differences relative to a closest family.
Given that features are static properties, such explanations provide \textit{no additional insights beyond a simple diff calculation}. 
We also observe that CADE's explainer~(\autoref{fig:cade_drift_exp}) \revOne{highlights} more features than those genuinely different, with only $25\%$ being accurate, which fails to narrow the analysis scope and even provides misleading information.

\subsection{Research Scope}

In this research, we employ active learning to mitigate Android malware concept drift, as depicted in \autoref{fig:active-learning}.
\tcolor{Cybersecurity teams deploy the malware classifier in a production environment online, while another system periodically evaluates its ability to make accurate decisions on incoming samples and take appropriate actions.}

This system consists of three key components~(formalized in ~\autoref{sec:formalization}).
The detector (\circlednuminline{1}) provides drift scores to indicate how much samples deviate from the existing distribution of labeled data. 
Detected samples, ranked and selected within a labeling budget, are then presented to malware analysts for review.
The explainer (\circlednuminline{2}) links detection decisions to semantically meaningful behaviors~(features in existing research~\cite{yang2021cade}), helping analysts understand the rationale behind the drift. 
Once the samples are labeled, the adaptor (\circlednuminline{3}) uses them to expand the original training data and dynamically update the classifier.

Recognizing that intra-class drift scenarios have been more extensively explored, our work primarily addresses the challenges associated with inter-class drifts, specifically those caused by previously unseen malware families. 
\tcolor{Our goal is to enhance detection and adaptation strategies for such drifts, enabling accurate malware classification in the dynamic environment.
By accurately identifying malware families, cybersecurity teams can take more targeted incident responses.}
Nevertheless, our proposed solution is not limited to inter-class drifts, as its generalizability to intra-class drifts will also be evaluated in~\autoref{eva:extend}.

%% file: method.tex
\section{Our \framework{} System}

We propose a system, named \framework{}, for drift detection and drift adaptation within the active learning framework.
In this section, we introduce design insights and technical details of the system.



\subsection{Design Insights}

\tcolor{To overcome the identified limitations and challenges, our core idea is to establish a \textit{bidirectional connection} between the classifier and the drift detector, while embedding high-level \textit{malware explanations} directly into the learning process.
As shown in~\autoref{fig:design_insight}, we design the drift detector using an autoencoder structure that incorporates two key features:
1)~aligning reconstructed samples with original samples in classifier predictions and 2)~embedding malware behavioral concepts in latent representations.
These features enable the drift detector to be informed by the \textit{classifier knowledge} while also allowing \textit{expert insights} to guide the classifier during adaptation, respectively addressing \textbf{L1} and \textbf{C1}, which directly impacts the accuracy of the updated classifier.} 

\tcolor{Specifically, we employ what we call \textit{model-sensitive concept learning} to train the drift detector. 
As shown in~\autoref{fig:embedding_insights}, the embedding space of the drift detector effectively separates different malware families with larger centroid distances, while drift samples are mapped to a clear outlier space. 
Furthermore, the alignment between the detector and classifier is evidenced by the classifier's activation embeddings: original and reconstructed samples from the same non-drift family cluster together, creating a well-defined decision boundary that accurately distinguishes drift samples. 
This consistency also suggests that the learned behavioral concepts can, in turn, enhance classifier retraining during adaptation.}

\tcolor{Following these designs, we achieve data autonomy~(solving \textbf{L2}) by relying on reconstructed samples to calculate drifting scores, \textit{eliminating the need for training data} during drift detection. To facilitate analysis on drift samples~(addressing \textbf{C2}) in adaptation, we adjust the drift explainer to generate explanations within the concept space, providing \textit{extended semantics beyond features} that align with human insights.}

\rcolor{
\bfnoindent{Key Notations}
Let \( \mathcal{X} \subseteq \mathbb{R}^{p \times q} \) denote the input feature space and \( \mathcal{Y} \) the output label space.
The classifier is represented as \( \mathbf{M} : \mathcal{X} \rightarrow \mathcal{P}(\mathcal{Y}) \), mapping features to label distributions. 
The latent embedding space is \( \mathcal{Z} \subseteq \mathbb{R}^N \), learned via an autoencoder \( f \), where an encoder \( f_{\text{enc}} \) maps \( \mathcal{X} \rightarrow \mathcal{Z} \) and a decoder \( f_{\text{dec}} \) reconstructs \( \mathcal{Z} \rightarrow \mathcal{X} \).  
We use \( \mathcal{D}_{\text{train}} \) and \( \mathcal{D}_{\text{test}} \) to refer to the training and test datasets, respectively. 
In CADE, the autoencoder is trained with a reconstruction loss \( \mathcal{L}_{\text{rec}} \) and a novel contrastive loss \( \mathcal{L}_{\text{sep}} \) to enforce separation between samples from different families.
In \framework{}, we enrich the latent space \( \mathcal{Z} \) into a malware concept space composed of an explicit concept space \( \mathcal{Z}_{\text{exp}} \) and an implicit concept space \( \mathcal{Z}_{\text{imp}} \). Built on this, we introduce a concept presence loss \( \mathcal{L}_{\text{pre}} \) to align embeddings with behavioral labels, and a reliability loss \( \mathcal{L}_{\text{rel}} \) to inject classifier knowledge for model-sensitive alignment.
We elaborate on their roles in enhancing the drift detection and adaptation process in the following sections.
}

\subsection{Concept-based Drift Detection}
\input{method_learn}

\subsection{Explanatory Drift Adaptation}
\input{method_adapt}




%% file: method_learn.tex

\bfnoindent{Model Sensitive Concept Learning} \label{sec:method_detection}
We explore the idea of concept learning to enhance the latent representations of the contrastive autoencoder. 
In machine learning, concept learning typically involves the task of inferring generalizable patterns (often framed as boolean-valued functions) from labeled examples to represent higher-level abstractions~\cite{jia2013visual, chiu2016design}. 
In our context, we envision that learning generalized malicious behavior concepts can guide the model in better identifying new malware families under drift. 
To this end, our goal is to integrate the supervised concept learning with the unsupervised contrastive learning. 



We define the unique latent space \(\mathcal{Z}\) in our method as the malware concept space, comprising the explicit concept space and the implicit concept space.
\rcolor{The explicit concept space \(\mathcal{Z}_{\text{exp}}\) is structured around a set of manually defined behaviors, where each element \( \mathbf{z}_{e}^{(i)} \) corresponds to a predefined malicious concept~(e.g., information stealing, remote control, and stealthy download).} 
This mirrors the classic concept learning, where boolean-valued functions are used to assign interpretable concepts to samples.
\rcolor{In contrast, the implicit concept space \( \mathcal{Z}_{\text{imp}} \) captures statistical patterns from the data and serves as an automatically learned latent representation.
It complements the explicit space by enriching the contrastive autoencoder and refining the contrastive loss \( \mathcal{L}_{\text{sep}} \) in two ways: 
1)~improving class separation when explicit behaviors overlap across malware families, and 2)~enabling concept space extensibility, such as accommodating benign behaviors in binary detection tasks.}

Built on the malware concept space, we introduce two key training requirements focused on concept handling. 
Firstly, we implement a concept presence loss for the precise detection of explicit concepts for each sample, which is defined as
\begin{equation}
    \mathcal{L}_{\text{pre}} = -\mathbf{m}_e \odot \left( \mathbf{l}_e \odot \log(\mathbf{p}_e) + (1 - \mathbf{l}_e) \odot \log(1 - \mathbf{p}_e) \right). 
    \label{equ:concept_presence}
\end{equation}
Here, $\odot$ denotes element-wise multiplication. The vectors \(\mathbf{m}_e\), \(\mathbf{l}_e\), and \(\mathbf{p}_e\) each have a length of $N_e$, representing the total number of explicit concepts.  
The elements \(\mathbf{m}_e^{(i)}\) and \(\mathbf{l}_e^{(i)}\) correspond to the valid label mask and the binary label for the i-th concept, respectively; and \(\mathbf{p}_e^{(i)}\) is calculated with \(g(\mathbf{z}_e^{(i)})\), which is the probability of the i-th explicit concept being present. 
This formula aggregates binary cross-entropy for all explicit concepts, incorporating the valid label mask \(\mathbf{m}_e\) to handle imprecise behavior labels.
This strategy addresses the challenge of behavior labeling in the malware domain, where unlike the image domain with direct human annotations, concept labels often originate from technical reports or dynamic analysis and can be uncertain or missing~\cite{smith2020mind}.

Our second innovation is the introduction of the concept reliability loss, endowing the detector with model sensitivity from the early training stage.
Our approach is based on the premise that a sample reconstructed from the concept space \( \mathcal{Z}_{\text{exp}} \cup \mathcal{Z}_{\text{imp}} \) should exhibit a probability distribution similar to the original sample when processed by the classifier \(\mathbf{M}\). 
The concept reliability loss is formally represented as
\begin{equation}
    \mathcal{L}_{\text{rel}} = -\sum_{i}^{C} \mathbf{M}(\mathbf{x})^{(i)} \log(\mathbf{M}(\hat{\mathbf{x}})^{(i)}).
    \label{equ:concept_reliability}
\end{equation}
In this equation, the probability distributions of the original sample \(\mathbf{x}\) serve as ``true labels'', and those of the reconstructed sample \(\hat{\mathbf{x}}\) are treated as predictions. 
This loss function effectively measures the divergence between the original and reconstructed sample distributions, ensuring that the model retains fidelity to the original concept representations even after reconstruction. 
Moreover, it implicitly measures the entropy-based uncertainty of the original distribution.

In summary, the training objective for our drift detector is centered on tuning the parameters of the autoencoder \(f\) and the concept presence function \(g\), minimizing
\begin{equation}
    \mathcal{L}(\mathcal{D}_{\text{train}}; \mathbf{l}, \mathbf{M}) = \lambda_0 \mathcal{L}_{\text{rec}} + \lambda_1 \mathcal{L}_{\text{sep}} + \lambda_2 \mathcal{L}_{\text{pre}} + \lambda_3 \mathcal{L}_{\text{rel}}, 
\end{equation}
where the traditional reconstruction loss, the concept-based contrastive loss, the concept presence loss, and the concept reliability loss are balanced by coefficients $\lambda_0$, $\lambda_1$, $\lambda_2$, and $\lambda_3$.

\bfnoindent{Data Autonomous Detector}
With the trained detector, we can leverage it to detect drifting samples for incoming test data \( \mathbf{x}_t \in \mathcal{D}_{\text{test}} \). 
Our design draws inspiration from a notable observation in the state-of-the-art intra-class detector HCC~(\autoref{sec:motivation_lesson}).
Considering the significant contribution of the cross-entropy-based pseudo loss to the performance, we focus on enhancing its contrastive-based element~(\autoref{equ:hcc_sampler}). 
We achieve this by maintaining the neighborhood function while substituting \( \hat{\mathcal{L}}_{hc} \) with \( \hat{\mathcal{L}}_{ce} \), leading to the introduction of the neighborhood cross-entropy (NCE) based pseudo loss, defined as
\begin{equation}
    \mathbf{D}_{\text{nce}}(\mathbf{x}_t; \mathcal{D}_{\text{train}}) = \hat{\mathcal{L}}_{\text{ce}}( \{\mathbf{x}_t\} \cup \mathcal{N}(\mathbf{x}_t; f, \mathcal{D}_{\text{train}})). 
    \label{equ:nce_detector}
\end{equation}
We observe that this adjustment can improve the HCC detector's performance to some extent, with the improvement in the average AUC achieves $6.11\%$ ~(see~\autoref{tab:intra-drift-nce}).


To comprehend the underlying principle, it's important to understand that NCE calculates uncertainty for nearest neighbors based on the assumption that a non-drifting sample should remain certain even if slightly perturbed.
This leads us to the samples reconstructed by the autoencoder, which essentially represent a constrained perturbation.
Consequently, the pseudo loss defined with our concept reliability loss (\autoref{equ:concept_reliability}) can align well with this principle.
The drift scoring function of our \textit{Concept Reliability Detector}~(CRD) is 
\begin{equation}
    u_d := \mathbf{D}_{\text{CRD}}(\mathbf{x}_t) =  \hat{\mathcal{L}}_{\text{ce}}(\mathbf{x}_t; \mathbf{M}) + \lambda_3 \hat{\mathcal{L}}_{\text{rel}}(\mathbf{x}_t, f(\mathbf{x}_t)),
    \label{equ:dream_detector}
\end{equation}
where the first item is the pseudo cross-entropy loss of the test sample in terms of the original classifier, and the second can be interpreted as the deviation in uncertainty after meaningful perturbations.
This approach offers advantages for two primary reasons: 
1)~the perturbation \( \hat{\mathbf{x}_t}=f(\mathbf{x}_t)\) is confined within the meaningful concept space, and the loss calculation considers the entropy of both the original sample and the perturbation; 
2)~the testing phase of the detector is specifically designed for data autonomy, ensuring its efficiency and operational independence from the training data.


%% file: method_adapt.tex
\begin{figure}[tb]
    \centering
    \includegraphics[width=\linewidth]{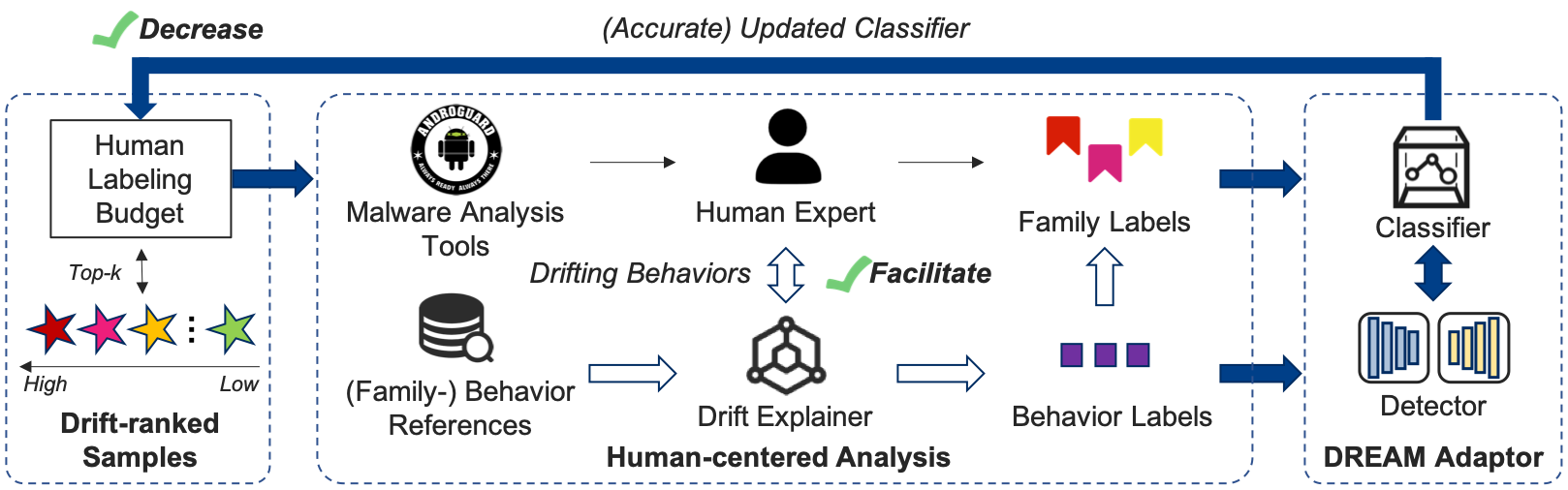}
    \caption{Human-in-the-loop strategies in \framework{}. \revOne{Bold blue arrows emphasize two core interaction flows: solid arrows denote system-level adaptation for efficient retraining with low budget, while hollow arrows represent human-centered behavioral explanation to assist labeling.}}
    \label{fig:dream_human}
\end{figure}

As shown in~\autoref{fig:dream_human}, \framework{} addresses the human-in-the-loop challenges with two strategies: 1)~a human-centered explanation mechanism to facilitate behavior labeling that can operate on the classifier, and 2)~a system-level behavior-involved adaptation algorithm to reduce the labeling budget required for maintaining classifier accuracy.

\bfnoindent{Explaining with Concept} \label{sec:method_explainer}
To implement the drift explainer, CADE introduces two key components: a perturbation function and a deviation function~(\(per\) and \(dev\) as in~\autoref{equ:explainer}).
Specifically, given a drifting sample \(\mathbf{x}_d\), it selects a reference sample \( \mathbf{x}_r \) from the training data by first identifying the closest class centroid \(y_r\), and then choosing the training sample whose latent representation is nearest to this centroid.
The perturbation process involves using the mask \( \mathbf{m} \in \mathbb{R}^{p \times q} \), resulting in the perturbed sample 
\begin{equation}
    \mathbf{x}_d^{\prime}:= \mathbf{x}_d \odot (1-\mathbf{m}) + \mathbf{x}_r \odot \mathbf{m} .
    \label{equ:feature_perturbation}
\end{equation}
The deviation is then quantified by calculating the distance between the latent representation of the perturbed sample, $\mathbf{z}_{d}^{\prime}$, and the nearest class centroid, represented as \( d( \mathbf{z}_{d}^{\prime}, \mathbf{c}_{y_r} ) \).

Motivated by this approach, our method focuses on generating concept-based explanations while leveraging the model-sensitive nature of our detector. 
First of all, in terms of the selection of the nearest class \(y_r\), we directly use \(\hat{y}\) which is the class predicted by the original classifier.
To create concept-based explanations, we introduce a concept-space mask \( \mathbf{m}_c \in \mathbb{R}^{N} \). 
This mask redefines the perturbation function within the concept space, while still allowing us to compute the original distance-based deviation using latent representations.
Further enhancing the deviation function, we incorporate the model-sensitive drifting scoring function. 
Therefore, the primary component of the explainer's optimization function is 
\begin{equation}
\begin{aligned}
    & d\left( \mathbf{z}_{d}^{\prime}, \mathbf{c}_{\hat{y}} \right) + \alpha_u \left( \hat{\mathcal{L}}_{\text{ce}}\left( \mathbf{x}_d^{\prime} \right) + \hat{\mathcal{L}}_{\text{rel}}\left( f\left(\mathbf{x}_r\right), \mathbf{x}_d^{\prime} \right) \right) , \\
    & \operatorname*{s.t.} \mathbf{z}_d^{\prime}:= \mathbf{z}_d \odot (1-\mathbf{m}_c) + \mathbf{z}_r \odot \mathbf{m}_c ,\ \mathbf{x}_d^{\prime} := f_{\text{dec}}(\mathbf{z}_d^{\prime}),
\end{aligned}
\label{equ:dream_explainer}
\end{equation}
where the drifting score is balanced by $\alpha_u$ and determined by the feature-space perturbed sample \(\mathbf{x}_d^{\prime}\) decoded from the perturbed concepts.
Specifically in the pseudo concept reliability loss $\hat{\mathcal{L}}_{\text{rel}}$, the sample reconstructed from the reference input, i.e., \(f(\mathbf{x}_r)\), acts as a proxy for the in-distribution output probabilities. 
Note that our method can still generate feature-space explanations if needed: $\mathbf{x}_d^{\prime}$ is created with feature-space operations as in~\autoref{equ:feature_perturbation}, so that pseudo concept reliability loss will be in its typical form, which is $\hat{\mathcal{L}}_{\text{rel}}(\mathbf{x}_d^{\prime}, f(\mathbf{x}_d^{\prime}))$. 





\bfnoindent{Concept-involved Adaptor}
Besides the drift explainer, our detector can serve as an explainer for in-distribution data in relation to the classifier. 
This is facilitated by the development of the explicit concept space. In this context, the outputs from the concept presence function $g$ are intrinsically linked to the malicious behaviors identified by the system.
When dealing with out-of-distribution samples, it is possible that these explanations may be inaccurate due to the evolving nature of the concepts they are based on. 
Nevertheless, the architecture of our detector is designed to allow malware analysts to utilize the drift explainer as a tool and refine these behavioral concepts. 
This adaptable design is instrumental in updating and enhancing the classifier’s performance in response to conceptual drifts.

We enable human analysts to provide feedback not only on the predicted labels of drifting samples but also on their behavioral explanations~(predicted concepts). 
For instance, in malware classification tasks, analysts might encounter a drifting sample identified as belonging to the \texttt{GhostCtrl} family, exhibiting malicious behaviors like \texttt{PrivacyStealing}, \texttt{SMSCALL}, \texttt{RemoteControl}, and \texttt{Ransom}. Feedback in this scenario can be formatted as: 
\begin{mdframed}[style=mystyle]
\begin{equation*}
\begin{split}
\exists mal \,\,\,\, &(\texttt{Family}(mal, \texttt{GhostCtrl})) \land (\texttt{Behaviors}(mal) \Leftrightarrow \\
             &\texttt{PrivacyStealing}(mal) \land \texttt{SMSCALL}(mal) \\
             &\land \texttt{RemoteControl}(mal) \land \texttt{Ransom}(mal) ).
\end{split}
\end{equation*}
\end{mdframed}

Utilizing the feedback provided, our drift adaptor operates by concurrently tuning the classifier and the detector.
Let $\Theta$ represent the parameters of the classifier's model, and $\Psi$ denote those of the detector.
The optimization problem, which aims to minimize the combined loss functions of the classifier and the detector, is formally defined as 
\begin{equation}
\begin{aligned}
    &\min_{\{\Theta, \Psi\}} \left( \mathcal{L}_{\text{ce}(\Theta)} + \mathcal{L}_{\text{det}(\Psi)} + \lambda_3 \mathcal{L}_{\text{rel}(\Theta, \Psi)} \right), \\
    &\operatorname{s.t. } \mathcal{L}_{\text{det}} = \lambda_0 \mathcal{L}_{\text{rec}(\Psi)} + \lambda_1 \mathcal{L}_{\text{sep}(\Psi)} + \lambda_2 \mathcal{L}_{\text{pre}(\Psi)}. \\
\end{aligned}
    \label{equ:concept_adaptor}
\end{equation}
The label feedback takes effect on the malware classification loss $\mathcal{L}_{\text{ce}}$ and the explanation feedback influences the concept presence loss $\mathcal{L}_{\text{pre}}$.
Contrasting with the detector's training process where $\Theta$ is fixed~(\autoref{equ:dream_detector}), during the adaptation phase, both $\Theta$ and $\Psi$ are subject to influence the concept reliability loss $\mathcal{L}_{\text{rel}}$. 
To address the complexities of joint parameter updating, we've implemented a dynamic learning rate schedule for the detector. 
Specifically, when $\mathcal{L}_{\text{det}}$ indicates that concept stability falls below a certain threshold, we reduce the learning rate using a scaling factor $\eta$. 
This reduction is based on the principle that stable concepts require less aggressive updates, promoting smoother model convergence and reducing the risk of overfitting.


%% file: evaluation.tex
\begin{figure*}[tb]
\centering
    \subfloat[Transcendent, \textsc{Drebin}]{%
        \includegraphics[width=0.24\linewidth]{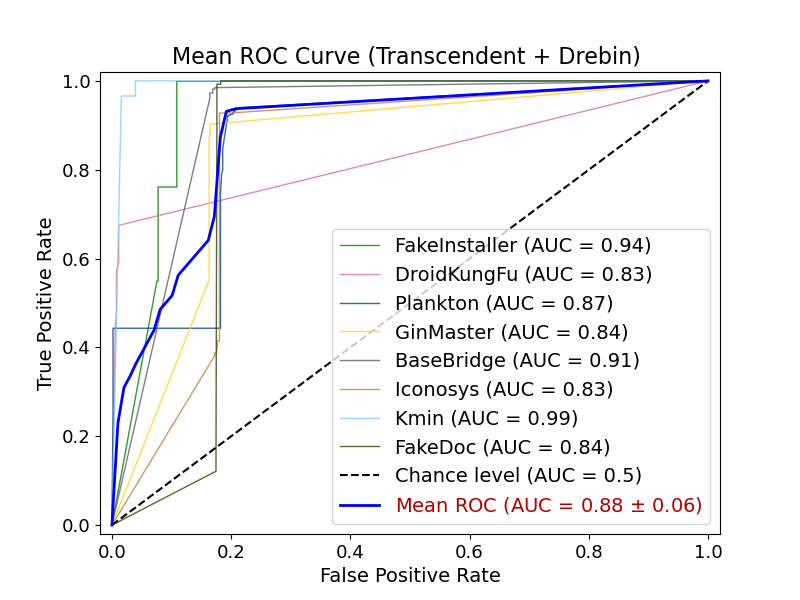}%
        \label{fig:a}}%
    \hfill
    \subfloat[CADE, \textsc{Drebin}]{%
        \includegraphics[width=0.24\linewidth]{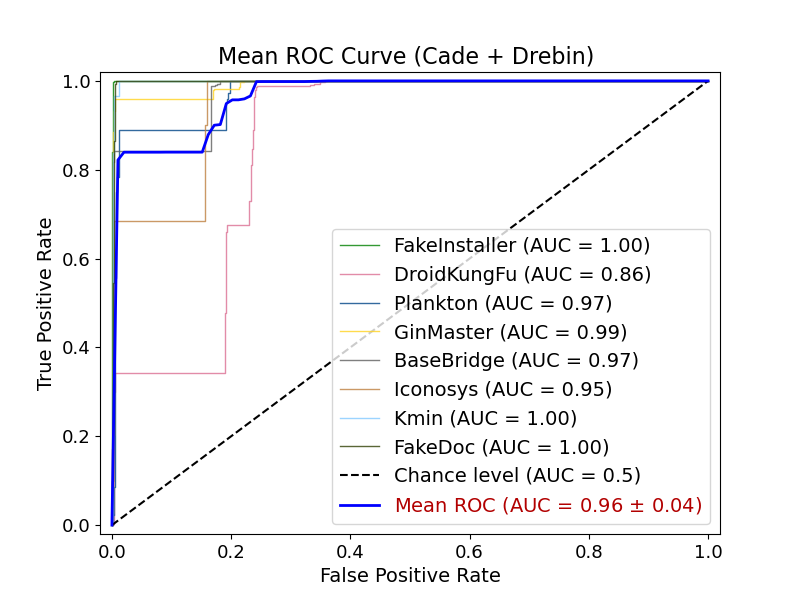}%
        \label{fig:b}}%
    \hfill
    \subfloat[Probability, \textsc{Drebin}]{%
        \includegraphics[width=0.24\linewidth]{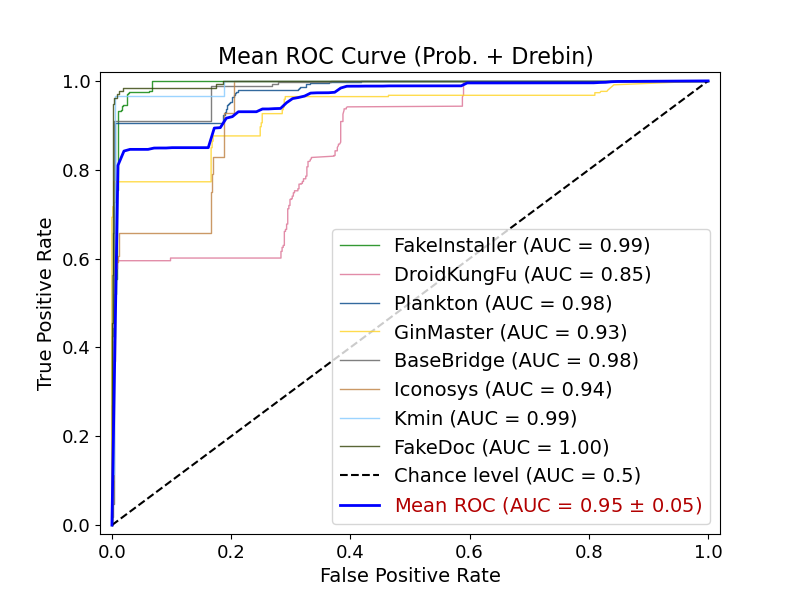}%
        \label{fig:c}}%
    \hfill
    \subfloat[(Ours) \framework{}, \textsc{Drebin}]{%
        \includegraphics[width=0.24\linewidth]{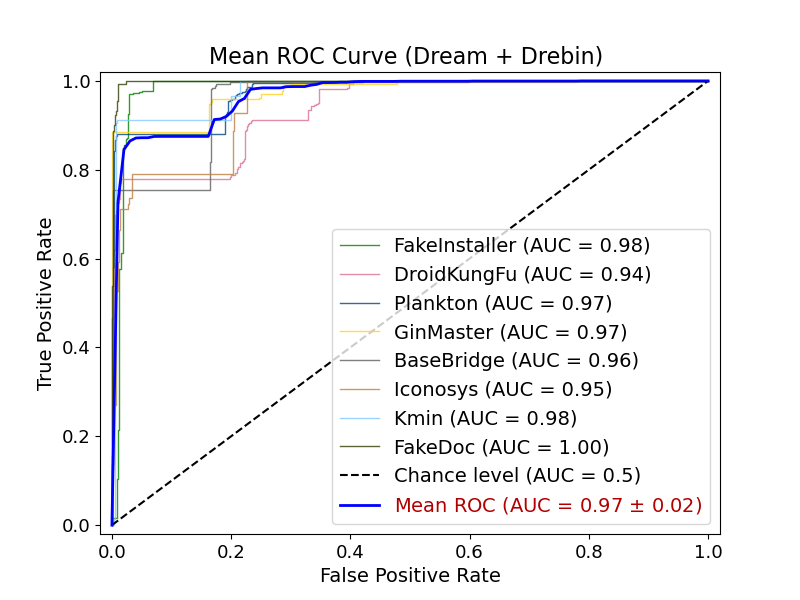}%
        \label{fig:d}}
    \vspace{-1em}

    \subfloat[Transcendent, \textsc{Mamadroid}]{%
        \includegraphics[width=0.24\linewidth]{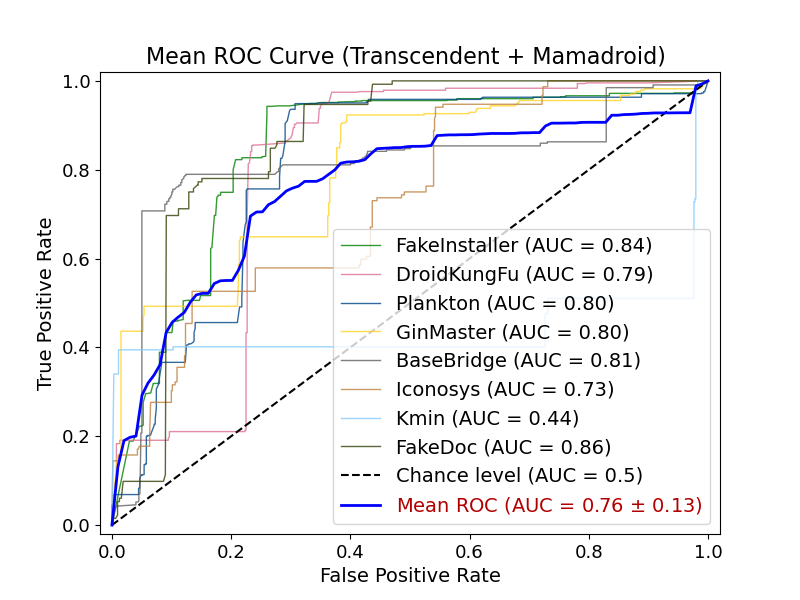}%
        \label{fig:e}}%
    \hfill
    \subfloat[CADE, \textsc{Mamadroid}]{%
        \includegraphics[width=0.24\linewidth]{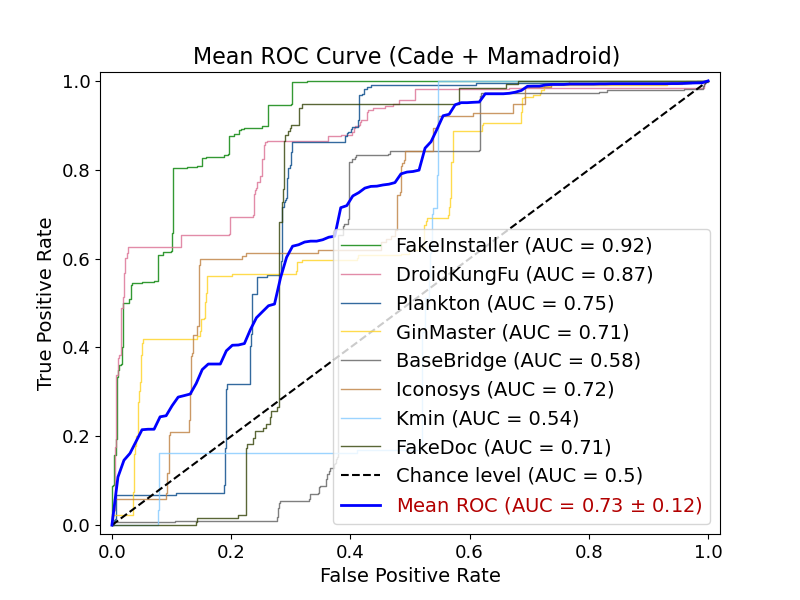}%
        \label{fig:f}}%
    \hfill
    \subfloat[Probability, \textsc{Mamadroid}]{%
        \includegraphics[width=0.24\linewidth]{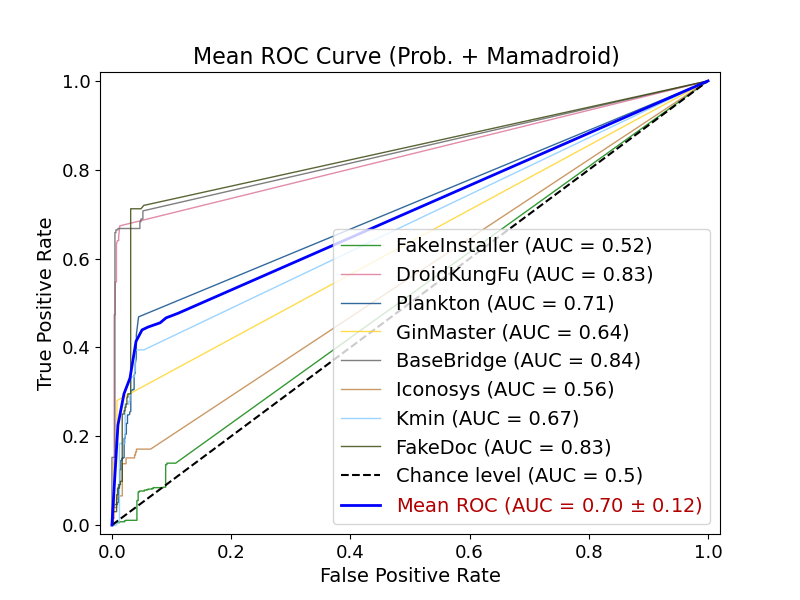}%
        \label{fig:g}}%
    \hfill
    \subfloat[(Ours) \framework{}, \textsc{Mamadroid}]{%
        \includegraphics[width=0.24\linewidth]{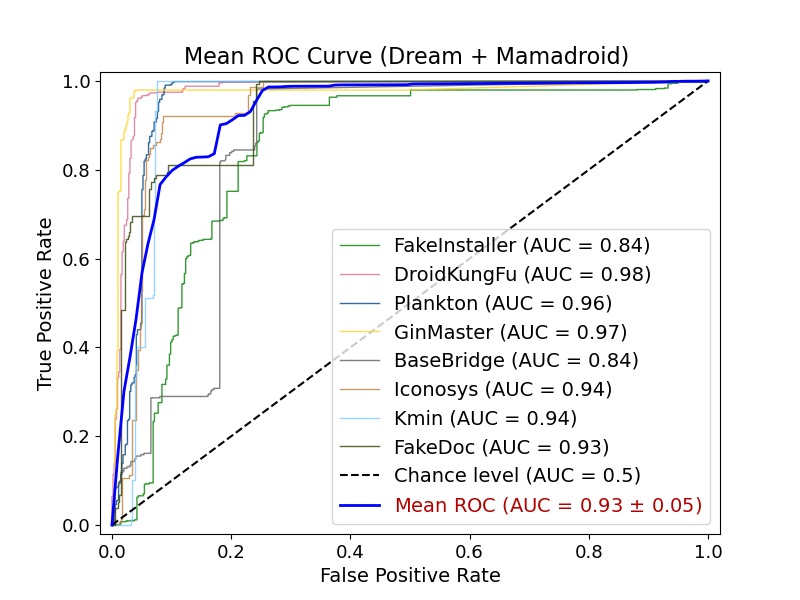}%
        \label{fig:h}}
    \vspace{-1em}

    \subfloat[Transcendent, \textsc{Damd}]{%
        \includegraphics[width=0.24\linewidth]{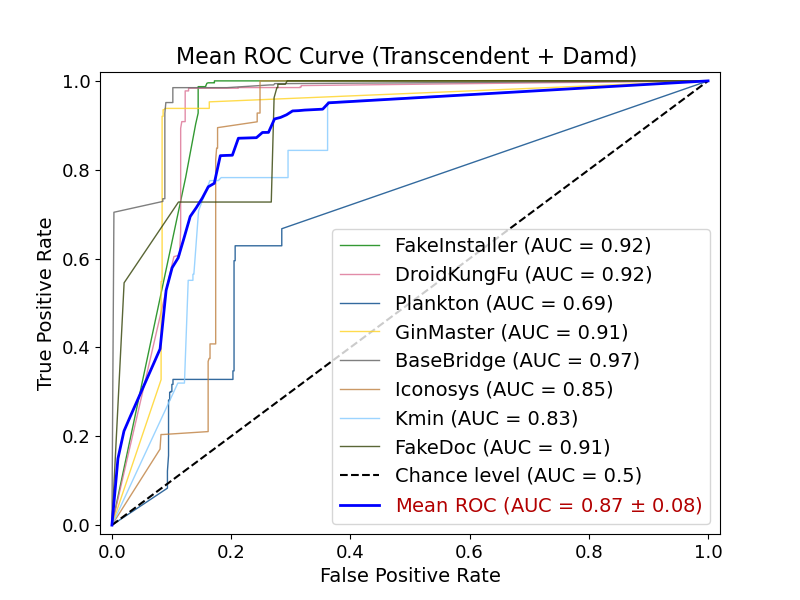}%
        \label{fig:i}}%
    \hfill
    \subfloat[CADE, \textsc{Damd}]{%
        \includegraphics[width=0.24\linewidth]{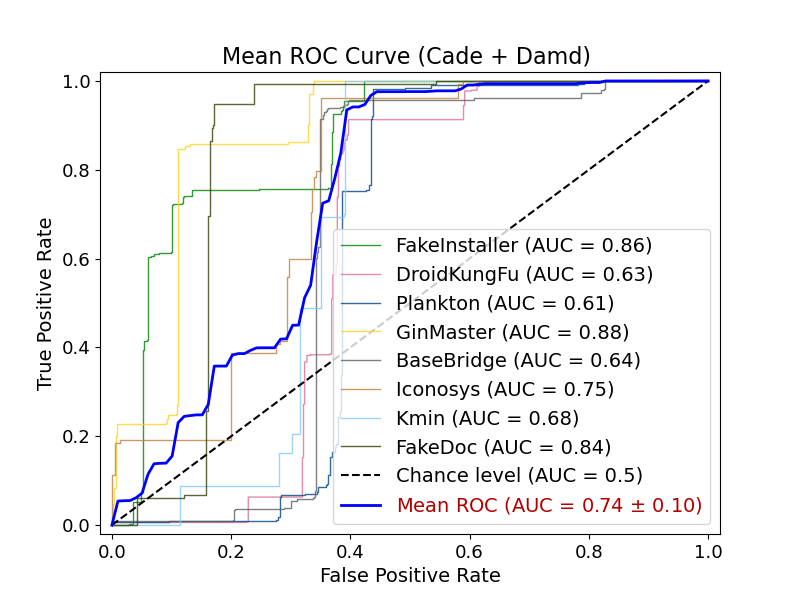}%
        \label{fig:j}}%
    \hfill
    \subfloat[Probability, \textsc{Damd}]{%
        \includegraphics[width=0.24\linewidth]{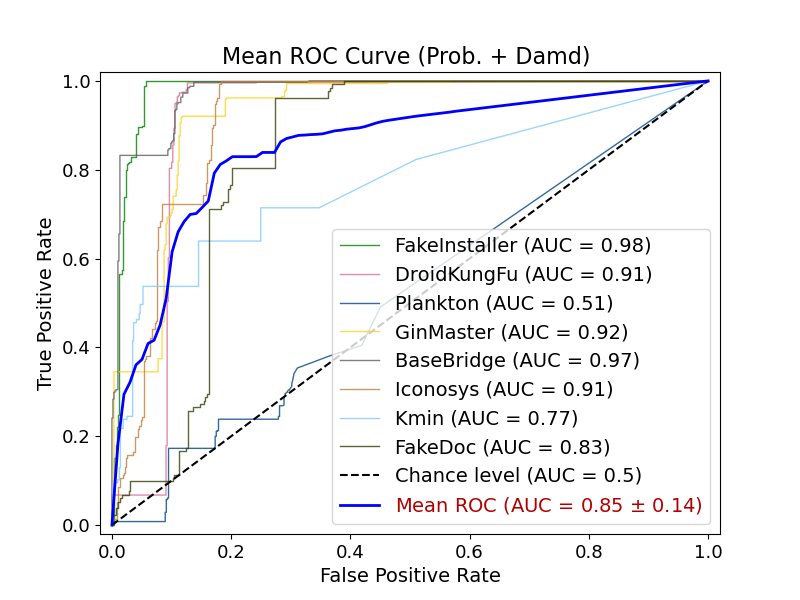}%
        \label{fig:k}}%
    \hfill
    \subfloat[(Ours) \framework{}, \textsc{Damd}]{%
        \includegraphics[width=0.24\linewidth]{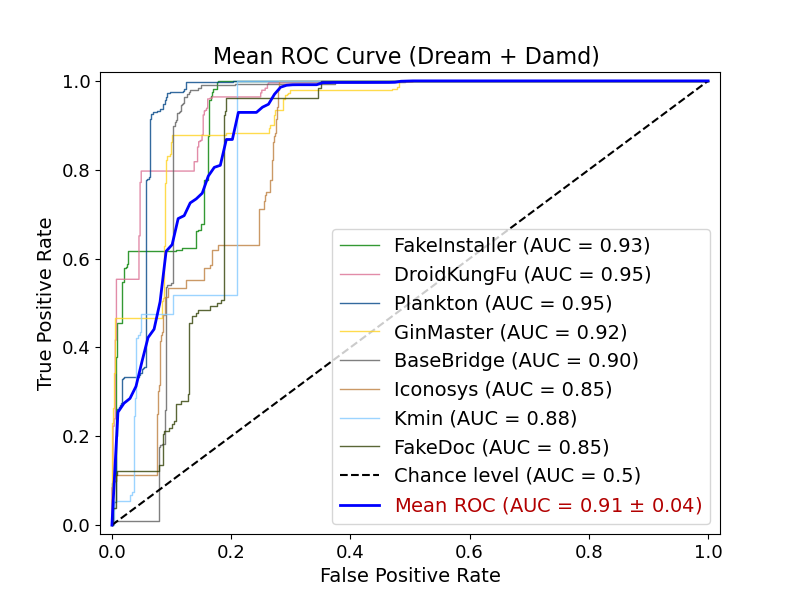}%
        \label{fig:l}}
    \vspace{-.8em}
    \caption{Evaluation of inter-class drift detection on Drebin dataset with three feature spaces. The first three columns are Transcendent, CADE, Probability, respectively, and our method is on the last column.} %
    \label{fig:multi-detection-basic}
\end{figure*}

This section presents a systematic evaluation of \framework{}, specifically focusing on inter-class drift scenarios, which are central to our research. 
We analyze the drift detector~(\autoref{sec:eva_detection}), the drift adaptor, and their joint effectiveness in updating the classifier~(\autoref{sec:eva_adaptation}). 
We also evaluate the drift explainer~(\autoref{sec:eva_explanation}) and perform a computational and human effort analysis of the system~(\autoref{sec:eva_human}).

\subsection{Experimental Setup} \label{sec:eva_setup}

\bfnoindent{Dataset}
We employ two malware datasets for malware family classification tasks, as illustrated in \autoref{tab:drebin_malradar}. 
Our first dataset is the well-known Drebin dataset \cite{arp2014drebin}. 
To capture more recent trends in the evolving malware landscape, we utilize the Malradar dataset \cite{wang2022malradar}. 
For each dataset, we select $8$ families with the same criteria used in the CADE paper, each comprising at least $100$ malware samples. 
The resulting datasets consist of $3,317$ and $2,589$ malware samples for Drebin and Malradar, respectively. 
Both datasets are well-labeled with respect to malware families. 
Malradar also provides behavioral labels derived from threat reports for each family, and we select $10$ distinct behaviors to form its behavioral labels: privacy information stealing~($\text{b}_0$), abusing SMS/CALL~($\text{b}_1$), remote control~($\text{b}_2$), bank/financial stealing~($\text{b}_3$), ransom~($\text{b}_4$), abusing accessibility~($\text{b}_5$), privilege escalation~($\text{b}_6$), stealthy download~($\text{b}_7$), aggressive advertising~($\text{b}_8$), and premium service~($\text{b}_9$).
To ensure consistency in behavioral analysis, we augment the Drebin dataset with these same behaviors. 
For more details about the datasets, please also refer to~\autoref{app:dataset}.

\bfnoindent{Classifier}
We use three deep learning-based malware classifiers, leveraging the features and models defined in previous works. 
These \revOne{classifiers} differ in data modality and model complexity. 
1)~\textsc{Drebin}~\cite{arp2014drebin} classifier utilizes eight feature sets representing binary vectors of predefined patterns, such as required permissions and suspicious API calls. 
The underlying model is a MLP~\cite{grosse2017adversarial}, configured with two hidden layers, sized at $100$ and $30$ neurons respectively.
2)~\textsc{Mamadroid}~\cite{mariconti2016mamadroid} classifier involves extracting API call pairs and abstracting them into package call pairs. It builds a Markov chain to model the transitions between packages, using the derived float vectors as features. 
Its MLP architecture includes hidden layers with dimensions of $1,000$ and $200$. 
3)~\textsc{Damd}~\cite{mclaughlin2017deep} classifier leverages the raw opcode sequences as features.
The sequence representation utilizes an embedding technique with a vocabulary size of $218$ tokens and an embedding dimension of $128$. 
Its underlying model is a CNN tailored for this task, featuring two convolutional layers, each with $64$ filters. 

\bfnoindent{Hold-out Strategy}
To evaluate inter-class drifts, where the drift labels are determined based on whether a malware family was unseen during training, we employ a commonly used hold-out strategy~\cite{yang2021cade, smith2020mind}.
To implement this, we first exclude samples of each malware family from the training set, reserving them solely for the testing phase. 
The remaining families are then divided into an 80:20 ratio for training and testing, adhering to a time-based separation criterion~\cite{ucci2019survey}. 
Since both datasets contain $8$ families, this strategy trains $8$ classifiers per dataset, matching the number of malware families.

\input{appendix/dataset_table}
\subsection{Drift Detection Performance} \label{sec:eva_detection}

\bfnoindent{Baseline and Metric}
For assessing our drift detector's performance in inter-class scenarios, we take into account the vanilla probability-based detector~(as depicted in~\autoref{equ:vanilla_detector}) and the innovative detectors from key studies summarized in~\autoref{tab:related}. The HCC detector is excluded from this evaluation for being incompatible with inter-class contexts.
To ensure a fair comparison, both the CADE detector and our detector, which each leverage an autoencoder model, are configured to share the same architecture across all features~(detailed in~\autoref{app:autoencoder}).
For the metric, we utilize the AUC calculated from the detector's drifting score output and the ground truth labels, which are determined by the held-out malware families during the training process.

\bfnoindent{Evaluation Results}
\autoref{fig:multi-detection-basic} and \autoref{fig:malradar_multi-detection-basic} depict the drift detection performance of our detector and the three baselines on the Drebin dataset and the Malradar dataset. 
Comparing the average AUC scores across different classifiers, we observe that \framework{} outperforms Transcendent, CADE, and Probability by 11.95\%, 15.64\%, and 12.4\%, respectively, on the Drebin dataset. Similarly, on the Malradar dataset, \framework{} shows an increase of 10.98\%, 8.33\%, and 14.7\%. 
In evaluating detection performance for different classifiers, we observed three key points. 
For the \textsc{Drebin} classifier, which is simpler and where most methods excel, the CADE detector performs comparably to our method on both datasets. 
However, for classifiers with more complex feature spaces, \revOne{such as the \textsc{Damd} feature within the MalRadar dataset}, CADE's effectiveness decreases by 26.5\% compared to the \textsc{Drebin} feature, while our method shows a 22.2\% advantage \revOne{over CADE}.
Regarding the \textsc{Mamadroid} classifier, it is less accurate in training compared to other classifiers. 
For this classifier, model-sensitive baselines such as Transcendent and Probability perform poorly. Our method, however, demonstrates stability and benefits from concept-based contrastive learning.
In the case of the \textsc{Damd} classifier, the most complex among the tested, the Transcendent detector shows a smaller performance decline relative to other baselines. 
This is likely due to its calibration process, which boosts accuracy in scenarios of slight classifier overfitting (evidenced by a test accuracy of only 0.94, even 2.97\% lower than that for the \textsc{Mamadroid} classifier). Although calibration proves beneficial, our method, with a focus on learning distance metrics, is more efficient, showing an average improvement of 6.62\% over the Transcendent detector.

\input{revision/adaptation_exp}

\subsection{Drift Adaptation Performance} \label{sec:eva_adaptation}

\bfnoindent{Baseline and Metric}
We compare our adaptation methods against commonly used techniques that retrain the classifier using selected samples and their annotated classification labels~\cite{jan2020throwing, chen2023continuous, yang2021cade}. 
\tcolor{For comprehensive comparison and ablation analysis, we integrate the baseline adaptor with previous drift detectors.
We categorize these four baseline adaptors into two groups, based on whether they use detectors that are data-autonomous or not.
In particular, the Probability baseline adaptor serves as a benchmark in scenarios where drift detection must be performed locally without training data and without components from \framework{}, allowing for a direct comparison under constraints similar to ours.
Note that although the CRD baseline employs drift detector within our system, it uses traditional retraining without incorporating the concept revision component in our adaptor.}

These experiments are conducted across a range of labeling budgets for active learning, specifically using \revOne{absolute numbers of labeled samples}—10, 20, 30, 40, and 100—with an emphasis on smaller budgets to minimize the need for extensive human annotation.
The effectiveness of each approach is quantified by measuring F1-scores and accuracy scores of the updated classifiers on the remaining test dataset. 
For inter-class adaptation, an important step is the modification of the classifier output to include new classes.
This modification involves randomly initializing the new output layer while preserving the learned weights in the existing layers, enabling continuous utilization of prior knowledge, and is consistently applied across all methods.


\bfnoindent{Evaluation Results}
\autoref{tab:adaptation_results} shows the drift adaptation results across different labeling budgets on all datasets and classifiers. 

\textit{1)~Bechmark Comparison.} 
Our method outperforms the benchmark in different settings without exception. 
Firstly, regarding performance improvements with different classifiers, we observed significant enhancements. On the Drebin dataset, the improvements of F1-score are 95.2\%, 82.7\%, and 22.6\% for \textsc{Drebin}, \textsc{Mamadroid}, and \textsc{Damd} classifiers, respectively. On the Malradar dataset, the improvements are  96.3\%, 11.6\%, and 39.2\% for the same classifiers.
Secondly, when considering varying human labeling budgets (10, 20, 30, 40, 100), the improvements of F1-score on the Drebin dataset are 238.1\%, 52.8\%, 21.1\%, 15.8\%, and 6.5\%, respectively, while being 98.7\%, 94.2\%, 23.3\%, 23.6\%, and 5.5\%  on the Malradar dataset for these respective budgets.
A key finding is that the smaller the budget, the greater the improvement. 
This suggests that in scenarios prioritizing human analysis, our method can significantly reduce labeling budgets. 
For example, achieving an accuracy score of 0.9 on the \textsc{Drebin} feature of the Drebin dataset requires labeling only 20 new samples with \framework{}, compared to 100 samples with the benchmark, reducing analysis cost by $80\%$.

\begin{table*}[tb]
  \centering
  \begin{tinytabularx}{.85\linewidth}{*{11}{C}}
    \toprule
    \textbf{Explainer} & \textbf{Metric} & \textbf{f0} & \textbf{f1} & \textbf{f2} & \textbf{f3} & \textbf{f4} & \textbf{f5} & \textbf{f6} & \textbf{f7} & \textbf{Avg.} \\ \hline
    \multirow{2}{*}{Random} & CBP & 0.239 & 0.145 & 0.131 & 0.070 & 0.093 & 0.191 & 0.225 & 0.057 & 0.144 \\
     & DRR & 0.212 & 0.175 & 0.218 & 0.153 & 0.201 & 0.202 & 0.324 & 0.316 & 0.225 \\ \hline
    \multirow{2}{*}{Dri-IG} & CBP & 0.154 & 0.217 & 0.166 & 0.246 & 0.103 & 0.194 & 0.299 & 0.208 & 0.198 \\
     & DRR & 0.134 & 0.281 & 0.287 & 0.464 & 0.263 & 0.263 & 0.469 & 0.463 & 0.328 \\ \hline
    \multirow{2}{*}{$\text{CADE}^\text{+}$} & CBP & 0.347 & 0.372 & 0.312 & 0.319 & 0.320 & 0.361 & 0.341 & 0.436 & 0.351 \\
     & DRR & 0.834 & 0.722 & 0.781 & 0.791 & 0.788 & 0.752 & 0.808 & \textbf{0.819} & 0.787 \\ \hline
    \multirow{2}{*}{\framework} 
     & CBP & \textbf{0.348} & \textbf{0.408} & \textbf{0.331} & \textbf{0.402} & \textbf{0.338} & \textbf{0.372} & \textbf{0.487} & \textbf{0.443} & \textbf{0.391} \\
     & DRR & \textbf{0.835} & \textbf{0.733} & \textbf{0.797} & \textbf{0.800} & \textbf{0.792} & \textbf{0.765} & \textbf{0.842} & \textbf{0.819} & \textbf{0.798} \\
    \bottomrule
  \end{tinytabularx}
\caption{Explanation evaluation results on the Drebin dataset with CBP and DRR metrics. The columns labeled f0 to f7 represent the held-out families, arranged in descending order by the number of samples in each family within the dataset.}
\label{tab:eva_concept_explainer}
\vspace{-10pt}
\end{table*}

\begin{figure*}
    \centering
    \subfloat[Reduced Budget]{%
    \includegraphics[width=0.3\linewidth]{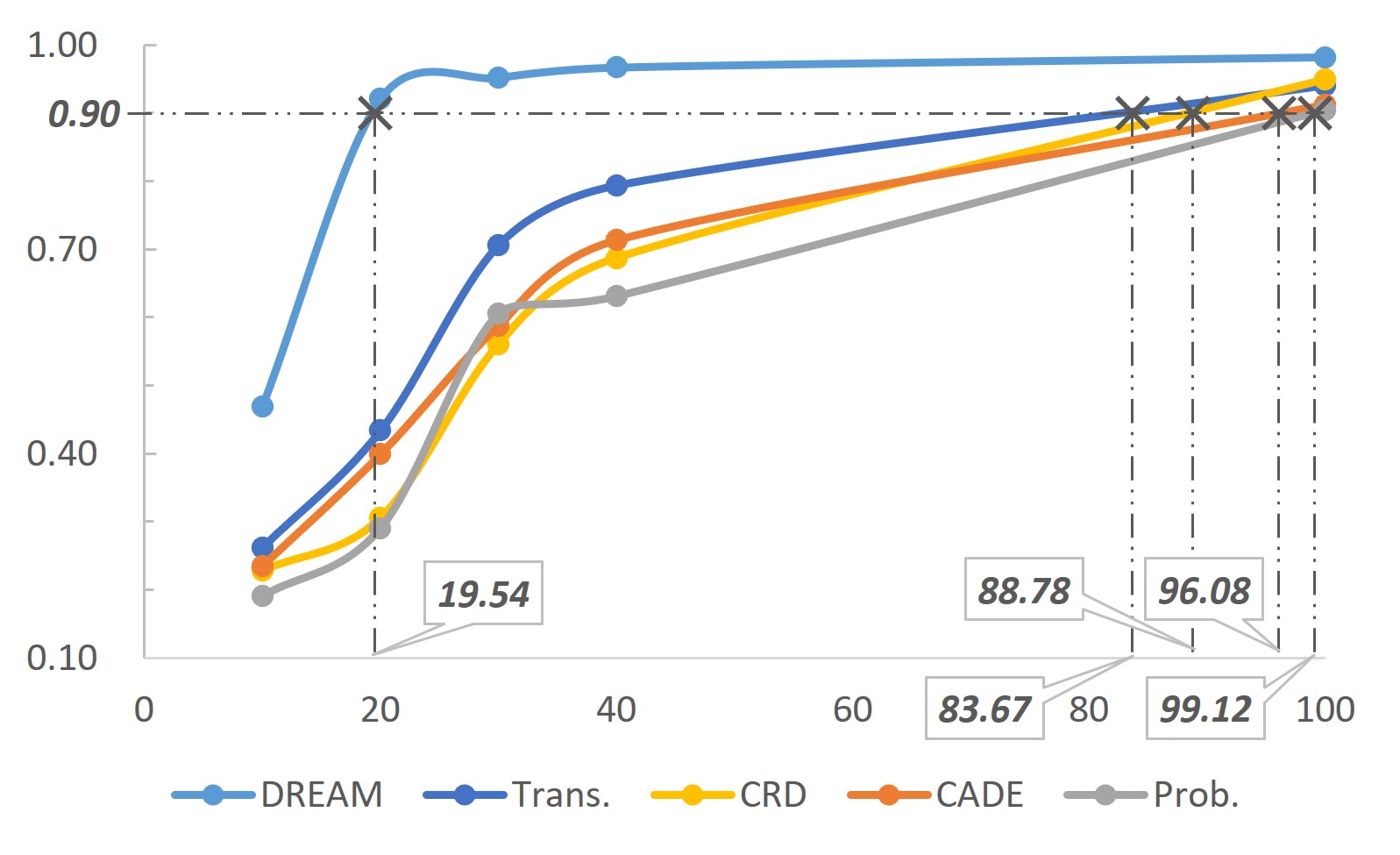}%
    \label{fig:reduced_budget}}%
    \hfill
    \subfloat[Useful Drift Explanation]{%
    \includegraphics[width=0.68\linewidth]{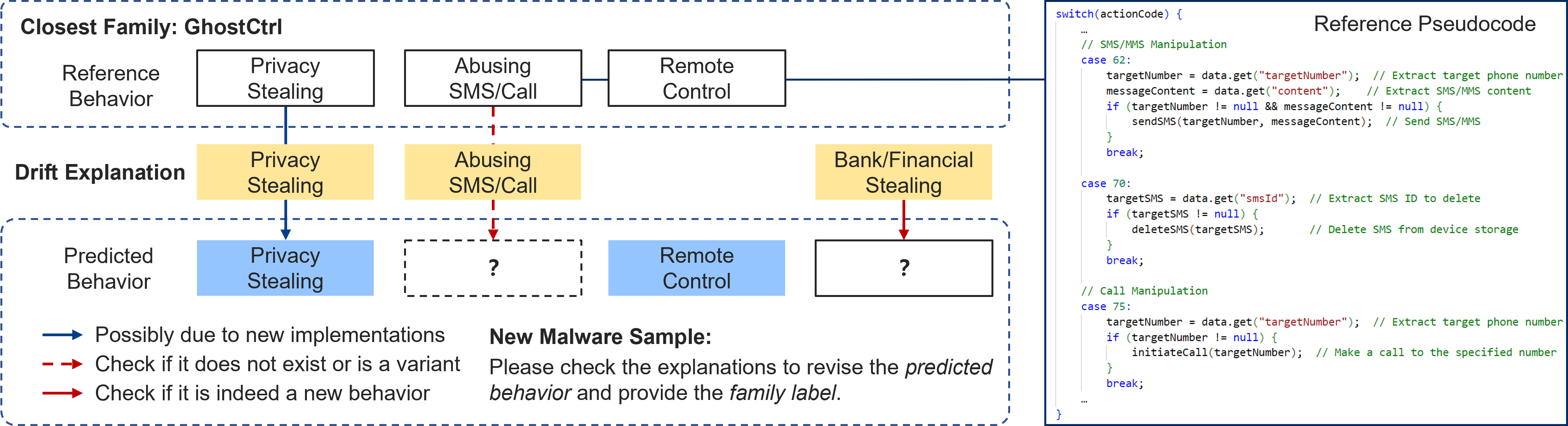}%
    \label{fig:drift_exp}}%
    \caption{\tcolor{Human effort analysis: demonstration of how \framework{} can (a)~reduce the labeling budget needed to maintain a certain accuracy level and (b)~facilitate the labeling process through explainability.}}
    \label{fig:human_effort_analysis}
\end{figure*}

\textit{\tcolor{2)~Existing Work Comparison.}} 
\tcolor{Compared with the two baselines that utilize data-dependent drift detectors from existing work, \framework{} maintains a clear advantage across the two datasets.
When evaluating across classifiers, \framework{} achieves F1-score improvements over Transcendent of 40.1\%, 34.7\%, and 31.1\%, respectively. The improvements over CADE are more substantial, at 89.0\%, 43.7\%, and 64.8\%.
Similarly, when comparing under varying annotation budgets, \framework{} outperforms Transcendent by 89.1\%, 47.2\%, 22.7\%, 13.1\%, and 4.5\%; compared with CADE, the improvements are 162.7\%, 77.5\%, 38.4\%, 30.7\%, and 20.0\%.}
\tcolor{Comparing the two baselines, we observe an interesting phenomenon: the Transcendent-based adaptor outperforms the CADE-based in over $83\%$ of cases, even if it identifies fewer true drift samples (detailed in \autoref{tab:adaptation details}). 
For instance, within the budget 10, even if Transcendent correctly detects only $2$ samples on average compared to CADE's $9.3$, the updated model using Transcendent still achieves a $0.25$ advantage in F1-score.
This performance can be attributed to Transcendent's focus on statistical decision boundaries rather than true malware patterns.
As a result, it inaccurately detects new families but improves retraining performance, especially for complex classifiers like those on \textsc{Damd}.
Transcendent even achieves slightly higher F1-score than \framework{} in two high-budget cases on this classifier within the Drebin dataset.
For the two specific instances, we combine Transcendent with our adaptor and improve its performance by $1.7\%$ and $1.4\%$. However, this combination requires training both our autoencoder and Transcendent's model sets, making it computationally expensive and limited to data-dependent detection scenarios.}


\textit{\tcolor{3)~Ablation Analysis.}}
\tcolor{Within the selected low budgets in adaptation, we observe that the Probability and CRD detectors are often comparable in accurately detecting samples from new families, outperforming existing detectors in $86.7\%$ cases.
Notably, our previous detection evaluation uses the AUC metric, showing that the CRD detector demonstrates an advantage in overall accuracy across different budgets. 
To clarify this difference, we threshold the number of samples that are truly drifting, finding that the CRD detector achieves an accuracy $14.56\%$ higher than the Probability detector.}
\tcolor{When the two detectors are integrated with the baseline adaptor, their performance remains similar, each succeeding in half of the specific cases.
However, when paired with \framework{} adaptor, we find that the Probability falls $12.8\%$ short of \framework{} in terms of F1-score of the updated classifier.
These findings suggest that the majority of performance gains in adaptation stem from our updating mechanism.
While the detected samples are crucial, their impact is highly specific to our system, as the primary value lies in how \framework{} utilizes them to update the classifier, with our detector incorporating behavioral concepts and actively participating in the process. 
}

\subsection{Drift Explanation Performance} \label{sec:eva_explanation}

\bfnoindent{Baseline and Metric}
We consider three baseline methods for drift explanation, adapted to generate both concept-level and traditional feature-level explanations: 
1)~a random baseline that randomly selects features or concepts as important; 
2)~a gradient-based explainer that utilizes Integrated Gradients (IG)~\cite{sundararajan2017axiomatic}, adapted here to analyze drifting scores derived from our detector. 
Originally designed to attribute a deep network's predictions to its input features, IG has demonstrated effectiveness in explicating supervised security applications~\cite{warnecke2020evaluating, he2023finer}.
In this context, it is tailored to focus on the drifting scores \revOne{derived} from model predictions, and the reference sample $\mathbf{x}_r$ serves as the baseline in its method; 
3)~the state-of-the-art CADE explainer used with the CADE detector, as described in \autoref{sec:method_explainer}. Since this method is not inherently designed for concept-space explanations, we adapt it to our detector for generating explanations in the concept space.

To evaluate the effectiveness of the drift explainers, we design two metrics. 
The first metric, named Cross Boundary P-value (CBP), assesses if explanations enable samples to cross the decision boundary, a key aspect in evaluating eXplainable AI~(XAI) methods~\cite{samek2016evaluating}.
In our context, CBP is quantified as the proportion of training set samples with higher drifting scores than the perturbed samples, formally represented as:
\begin{equation}
    \frac{|\{\alpha \in \mathcal{D}_{\text{train}}[\hat{y}_i] : u_d(\alpha; \mathbf{M}) \geq u_d(\mathbf{x}_d^{\prime}; \mathbf{M})\}|}{|\mathcal{D}_{\text{train}}[\hat{y}_i]|}.
\end{equation}
The second metric, as suggested in existing research~\cite{yang2021cade}, focuses on the distance to the reference sample in the detector's latent space after perturbation. 
We employ the Distance Reduction Rate~(DRR) for this purpose, which measures the ratio of the reduced distance to the original distance between the drifting sample and the reference sample.
It is worth mentioning that CBP is our primary metric of interest, as it directly correlates with classifier results, whereas DRR serves more as a supplementary measure in the latent space.

For a balanced comparison, particularly against the first two baseline methods that only yield importance scores, we align the number of altered features or concepts with those pinpointed by our techniques. 
These experiments are conducted using the \textsc{Drebin} feature of the Drebin dataset, maintaining consistency with the CADE paper, where the performance of its drift detector is established.

\bfnoindent{Evaluation Results}
\autoref{tab:eva_concept_explainer} illustrates the evaluation results of concept-based explanations generated by different methods. 
Comparing to three baseline explainers, i.e., Random, Dri-IG and $\text{CADE}^\text{+}$, \framework{} surpasses them by 172.1\%, 97.2\% and 11.5\% on the CBP metric, and by 254.2\%, 143.3\% and 1.4\% on the DRR metric. 
Notably, $\text{CADE}^\text{+}$ is included in this comparison as it utilizes our detector for concept-based explanations, extending beyond its original method's capabilities. 
To examine the efficacy in areas typically addressed by common methods, \autoref{tab:feature_explainer_eva} presents the outcomes for feature-level explanation evaluations. 
Here, our method, despite not being primarily designed for feature-level explanations, outperforms all baselines, including the state-of-the-art CADE explainer, benefiting from our method's sensitive capture of deviations. 
Combining the results from both tables, our explainer yields more substantial improvements in concept space than in feature space. 
For instance, while a 45.2\% increase in mean CBP over Dri-IG is shown at the feature level, a remarkable 97.2\% increase is illustrated at the concept level, highlighting the effectiveness of our design in concept space.

\input{revision/computation_exp}












%% file: appendix/dataset_table.tex
\begin{table}[tb]
    \centering
    \begin{smalltabularx}{\linewidth}{p{1.3cm}<{\centering}p{1.2cm}<{\centering}p{2.5cm}p{2cm}<{\centering}}
    \toprule
     \textbf{Family} & \textbf{\# sample} & \textbf{(\#) behavior} & \textbf{Time}\\ 
    \hline
    $\text{FakeInstaller}$ & 925 & $\text{(5) } \text{b}_0\text{, }\text{b}_1\text{, }\text{b}_7\text{, }\text{b}_8\text{, }\text{b}_9$ & $\text{2011-2012}$ \\
    \hline
    $\text{DroidKungFu}$ & 667 & $\text{(3) }\text{b}_0\text{, }\text{b}_2\text{, }\text{b}_6$  & $\text{2011-2012}$ \\  
    \hline
    $\text{Plankton}$ & 625 & $\text{(4) }\text{b}_0\text{, }\text{b}_2\text{, }\text{b}_7\text{, }\text{b}_8$ & $\text{2011-2012}$  \\ 
    \hline
    $\text{GingerMaster}$ & 339 & $\text{(2) }\text{b}_0\text{, }\text{b}_6$ & $\text{2011-2012}$\\
    \hline
    $\text{BaseBridge}$ & 330 & $\text{(4) }\text{b}_0\text{, }\text{b}_1\text{, }\text{b}_6\text{, }\text{b}_9$ & $\text{2010-2011}$ \\ 
    \hline
    $\text{Iconosys}$ & 152 & $\text{(2) }\text{b}_0\text{, }\text{b}_8$ & $\text{2010-2011}$ \\
    \hline
    $\text{Kmin}$ & 147 & $\text{(1) }\text{b}_0$ & $\text{2010-2012}$ \\  
    \hline
    $\text{FakeDoc}$ & 132 & $\text{(1) }\text{b}_4$ & $\text{2011-2012}$\\  
    \midrule
    \end{smalltabularx}

    \begin{smalltabularx}{\linewidth}{p{1.3cm}<{\centering}p{1.2cm}<{\centering}p{2.5cm}p{2cm}<{\centering}}
    \toprule
    $\text{RuMMS}$ & 796 & $\text{(4) }\text{b}_0\text{, }\text{b}_1\text{, }\text{b}_2\text{, }\text{b}_3$ & $\text{2016-2018}$\\
    \hline
    $\text{Xavier}$ & 589 & $\text{(4) }\text{b}_0\text{, }\text{b}_2\text{, }\text{b}_7\text{, }\text{b}_8$ & $\text{2016-2021}$\\  
    \hline
    $\text{LIBSKIN}$ & 290 &   $\text{(4) }\text{b}_0\text{, }\text{b}_1\text{, }\text{b}_2\text{, }\text{b}_6\text{, }\text{b}_7\text{, }\text{b}_8$ & $\text{2015-2021}$  \\ 
    \hline
    $\text{HiddenAd}$ & 289 & $\text{(2) }\text{b}_0\text{, }\text{b}_8$ & $\text{2017-2021}$ \\
    \hline
    $\text{MilkyDoor}$ & 210 &  $\text{(2) }\text{b}_0\text{, }\text{b}_2$ & $\text{2016-2020}$  \\ 
    \hline
    $\text{GhostClicker}$ &182 & $\text{(4) }\text{b}_0\text{, }\text{b}_2\text{, }\text{b}_6\text{, }\text{b}_8$ & $\text{2016-2020}$  \\
    \hline
    $\text{EventBot}$ & 124 &  $\text{(5) }\text{b}_0\text{, }\text{b}_1\text{, }\text{b}_2\text{, }\text{b}_3\text{, }\text{b}_5$ & $\text{2020}$ \\ 
    \hline
    $\text{GhostCtrl}$ & 109 & $\text{(3) }\text{b}_0\text{, }\text{b}_1\text{, }\text{b}_2$ & $\text{2016-2020}$  \\  
    \bottomrule
    \end{smalltabularx}
    \caption{Drebin~(top) and Malradar~(bottom) datasets. In the third column, the number of behaviors is in parentheses, followed by their indices.} 
    \label{tab:drebin_malradar}
    \vspace{-18pt}
\end{table}

%% file: revision/adaptation_exp.tex
\begin{table*}[tb]
    \begin{tinytabularx}{\linewidth}{>{\centering\arraybackslash}p{0.9cm} >{\centering\arraybackslash}p{0.7cm} >{\centering\arraybackslash}p{0.7cm}|CCCC|l|CCCC|l}
    \toprule
    \multirow{3}{*}{\textbf{Feature}} & \multirow{3}{*}{\textbf{Budget}} & \multirow{3}{*}{\textbf{\shortstack{Metric}}} &
    \multicolumn{5}{c|}{\textbf{Drebin Dataset}} &
    \multicolumn{5}{c}{\textbf{MalRadar Dataset}} \\
    \cmidrule(lr){4-8} \cmidrule(lr){9-13}
    &&& \multicolumn{4}{c|}{\textbf{Baseline Adaptor}} &
    \multirow{2}{*}{\textbf{\framework} \textbf{Adaptor}} &
    \multicolumn{4}{c|}{\textbf{Baseline Adaptor}} &
    \multirow{2}{*}{\textbf{\framework} \textbf{Adaptor}} \\
    \cmidrule(lr){4-7} \cmidrule(lr){9-12}
    &&& \footnotesize{Trans.} & \footnotesize{CADE} & \footnotesize{Prob.} & \footnotesize{CRD} &  & \footnotesize{Trans.} & \footnotesize{CADE} & \footnotesize{Prob.} & \footnotesize{CRD} &  \\
    \hline
    
    \multirow{10}{*}{\textsc{Drebin}}& \multirow{2}{*}{10}& F1& \textbf{0.416} & {0.166}& 0.173&\textbf{0.449}  
    & \textbf{0.805} \perc{93.33} \perc{79.09}
    & \textbf{0.262} & {0.235}& 0.191& {\textbf{0.228}} &\textbf{0.469} \perc{78.94} \perc{105.84} \\
    && Acc.& \textbf{0.407} & {0.172}& 0.168& {\textbf{0.418}} & \textbf{0.778} \perc{91.36} \perc{86.33} & \textbf{0.295} & {0.283}& 0.207& {\textbf{0.265}} & \textbf{0.525} \perc{78.04} \perc{97.81}    \\ \hhline{~------------} 
    
    & \multirow{2}{*}{20}& F1  & \textbf{0.703} & {0.407}& 0.528& {\textbf{0.562}} & \textbf{0.928} \perc{31.85} \perc{64.90} & \textbf{0.434} & {0.399}& 0.291& {\textbf{0.305}} & \textbf{0.921} \perc{112.30} \perc{201.64}   \\
    && Acc.& \textbf{0.677} & {0.402}& 0.497& {\textbf{0.531}} & \textbf{0.905} \perc{33.67} \perc{70.30} & 0.455& {\textbf{0.457}} & 0.336& {\textbf{0.349}} &\textbf{0.913}  \perc{99.95} \perc{161.26}   \\ \hhline{~------------} 
    
    & \multirow{2}{*}{30}& F1  & \textbf{0.852} & {0.749}& \textbf{0.785} & {0.717}& \textbf{0.961} \perc{12.72} \perc{22.34} & \textbf{0.706} & {0.587}& \textbf{0.605} & {0.560}& \textbf{0.952} \perc{34.85} \perc{57.23}    \\
    && Acc.& \textbf{0.834} & {0.712}& \textbf{0.744} & {0.697}& \textbf{0.952} \perc{14.14} \perc{27.95} & \textbf{0.726} & {0.622}& \textbf{0.629} & {0.585}& \textbf{0.948} \perc{30.69} \perc{50.76}    \\ \hhline{~------------} 
    
    & \multirow{2}{*}{40}& F1  & \textbf{0.890} & {0.856}& \textbf{0.886} & {0.781}&  \textbf{0.956} \percplus{7.42} \percplus{7.88}& \textbf{0.793} & {0.714}& 0.631& {\textbf{0.687}} & \textbf{0.967} \perc{21.87} \perc{40.73}    \\
    && Acc.& \textbf{0.878} & {0.810}& \textbf{0.849} & {0.755}& \textbf{0.943} \percplus{7.40} \perc{11.04} & \textbf{0.808} & {0.740}& 0.657& {\textbf{0.703}} & \textbf{0.964} \perc{19.29} \perc{37.12}    \\ \hhline{~------------} 
    
    & \multirow{2}{*}{100}    & F1  & 0.942& {\textbf{0.965}} & 0.923& {\textbf{0.937}} & \textbf{0.975} \percplus{1.10} \percplus{4.13}& \textbf{0.940} & {0.913}& 0.904& {\textbf{0.949}} &\textbf{0.982} \percplus{4.45} \percplus{3.49}\\
    && Acc.& 0.946& {\textbf{0.955}} & 0.901& \textbf{0.927} & \textbf{0.973} \percplus{1.87} \percplus{4.88}& \textbf{0.931} & {0.901}& 0.892& {\textbf{0.945}} & \textbf{0.981} \percplus{5.43} \percplus{3.83}\\ \hline

    \multirow{10}{*}{\makecell{\textsc{Mama-} \\ \textsc{droid}}} & \multirow{2}{*}{10}& F1  & 0.294& \textbf{0.313} & 0.169& \textbf{0.299} & \textbf{0.648} \perc{106.93} \perc{116.60}& \textbf{0.445} & 0.377& \textbf{0.476} & 0.474& \textbf{0.651} \perc{46.25} \perc{36.80}    \\
    &  & Acc. & 0.264 & \textbf{0.311} & 0.156 & 0.300 & \textbf{0.650} \perc{108.64} \perc{116.53} & 0.300 & \textbf{0.413} & 0.514 & \textbf{0.520} & \textbf{0.670} \perc{36.89} \perc{28.75} \\ \hhline{~------------} 
    
    & \multirow{2}{*}{20}& F1 & 0.443& {\textbf{0.490}} & \textbf{0.449} & {0.428}& \textbf{0.740} \perc{51.02} \perc{64.92} & \textbf{0.571} & {0.483}& 0.623& {\textbf{0.629}} & \textbf{0.718}  \perc{25.79} \perc{14.21}    \\
    && Acc.& 0.411& {\textbf{0.474}} & \textbf{0.435} & {0.426}& \textbf{0.736} \perc{55.16} \perc{69.32} & \textbf{0.602} & {0.519}& 0.626& {\textbf{0.654}} & \textbf{0.725} \perc{20.38} \perc{10.97}    \\ \hhline{~------------}
    
    & \multirow{2}{*}{30}& F1  & \textbf{0.610}& 0.582& 0.601& {\textbf{0.614}} & \textbf{0.789} \perc{29.37} \perc{28.59} & \textbf{0.615} & {0.594}& \textbf{0.718} & {0.669}& \textbf{0.734} \perc{19.50} \percplus{2.31}\\
    && Acc.& \textbf{0.573}& 0.566& 0.579& {\textbf{0.603}} & \textbf{0.780} \perc{36.05} \perc{29.28} & \textbf{0.641} & {0.623}& \textbf{0.722} & {0.687}& \textbf{0.737} \perc{14.88} \percplus{2.09}\\ \hhline{~------------} 
    
    & \multirow{2}{*}{40}& F1  & \textbf{0.712} & {0.587}& 0.630& {\textbf{0.677}} & \textbf{0.808} \perc{13.51} \perc{19.34} & \textbf{0.672} & {0.653}& \textbf{0.756} & {0.684}&\textbf{0.766}   \perc{14.00} \percplus{1.33}\\
    && Acc.& \textbf{0.677} & {0.590}& 0.621& {\textbf{0.670}} & \textbf{0.794} \perc{17.21} \perc{18.51} & \textbf{0.681} & {0.666}& \textbf{0.747} & {0.698}& \textbf{0.759} \perc{11.37} \percplus{1.51}\\ \hhline{~------------} 
    
    & \multirow{2}{*}{100}    & F1  & \textbf{0.827} & {0.642}& \textbf{0.814} & {0.807}& \textbf{0.864} \percplus{4.47} \percplus{6.22}& \textbf{0.785} & {0.769}& 0.821& {\textbf{0.830}} & \textbf{0.839}  \percplus{6.88} \percplus{1.10}\\
    && Acc.& \textbf{0.797}& {0.652}& \textbf{0.805} & {0.793}&\textbf{0.853} \percplus{6.97} \percplus{6.00}& \textbf{0.781} & {0.767}& 0.811& {\textbf{0.820}} & \textbf{0.829}  \percplus{6.12} \percplus{1.14}\\ \hline

    \multirow{10}{*}{\textsc{Damd}}& \multirow{2}{*}{10}& F1  & \textbf{0.510} & {0.339}& \textbf{0.442} & {0.363}& \textbf{0.738} \perc{44.60} \perc{66.91} & \textbf{0.209} & {0.179}& \textbf{0.246} & {0.227}& \textbf{0.525} \perc{151.16} \perc{113.68}   \\
    && Acc.& \textbf{0.437} & {0.360}& \textbf{0.422} & {0.353}& \textbf{0.702} \perc{60.62} \perc{66.38} & \textbf{0.263} & {0.231}& \textbf{0.303} & {0.294}& \textbf{0.556}  \perc{111.10} \perc{83.58}    \\ \hhline{~------------} 
    
    & \multirow{2}{*}{20}& F1  & \textbf{0.801} & 0.582& \textbf{0.711} & 0.659& \textbf{0.837} \percplus{4.50} \perc{17.77} & \textbf{0.482} & {0.420}& 0.454& {\textbf{0.495}} &\textbf{0.683} \perc{41.71} \perc{37.86}    \\
    && Acc.& \textbf{0.767} & 0.599& \textbf{0.684} & 0.640& \textbf{0.820} \percplus{6.98} \perc{19.92} & \textbf{0.535} & 0.474& 0.497& \textbf{0.538} & \textbf{0.707} \perc{32.16} \perc{31.27}    \\ \hhline{~------------} 
    
   & \multirow{2}{*}{30}& F1  & \textbf{0.840} & {0.610}& \textbf{0.775} & {0.687}& \textbf{0.849} \percplus{1.07}  \percplus{9.51}& \textbf{0.541} & {0.530}& \textbf{0.681} & {0.550}& \textbf{0.750} \perc{38.52} \perc{10.18}    \\
    && Acc.& \textbf{0.804} & {0.631}& \textbf{0.756} & {0.687}&  \textbf{0.844} \percplus{4.96} \perc{11.66} & \textbf{0.600} & {0.571}& \textbf{0.697} & {0.605}&\textbf{0.751} \perc{25.29} \percplus{7.86}\\ \hhline{~------------} 
    
    & \multirow{2}{*}{40}& F1  & \textbf{0.910} & {0.590}& \textbf{0.778} & {0.714}& \textbf{0.867} \percplus{-4.74} \perc{11.32} & \textbf{0.578} & {0.541}& \textbf{0.630} & {\textbf{0.630}} &   \textbf{0.732} \perc{26.66} \perc{16.11}    \\
   && Acc.& \textbf{0.881} & {0.628}& \textbf{0.766} & {0.697}& \textbf{0.863} \percplus{-2.05} \perc{12.71} & \textbf{0.626} & {0.595}& \textbf{0.669} & {0.655}& \textbf{0.749} \perc{19.69} \perc{11.92}    \\ \hhline{~------------} 
   
    & \multirow{2}{*}{100}    & F1  & \textbf{0.955} & {0.652}& 0.888& {\textbf{0.904}} & \textbf{0.955} \percplus{-0.01} \percplus{5.63}& \textbf{0.753} & {0.669}& 0.768& {\textbf{0.770}} &\textbf{0.811} \percplus{7.64} \percplus{5.29}\\
    && Acc.& \textbf{0.941} & {0.654}& 0.872& {\textbf{0.875}} &\textbf{0.941} \percplus{0.03} \percplus{7.55}& \textbf{0.761} & {0.697}& \textbf{0.778} & {0.777}& \textbf{0.809}    \percplus{6.29} \percplus{4.04}\\
    
    \bottomrule
    \end{tinytabularx}
    \caption{Drift adaptation results on two malware datasets across three feature sets. \tcolor{We integrate the baseline adaptor with different drift detectors and categorize them into two groups based on their applicable scenarios: those \revOne{that} require access to the original training data for drift detection (Transcendent, CADE) and those \revOne{that} do not (Probability, CRD). Baselines with the best performance in each scenario are highlighted in bold, and our respective improvement ratios are reported in blue.}} 
    \label{tab:adaptation_results}
    \vspace{-15pt}
\end{table*}

%% file: revision/computation_exp.tex
\subsection{\tcolor{Computational \& Human Effort Analysis}} \label{sec:eva_human}

\tcolor{The overhead of \framework{} comprises three primary components: 1)~computational investment required for detecting drift samples from unlabeled data, 2)~computational cost involved in tuning the model during drift adaptation, and 3)~human effort required to generate correct labels for accurate adaptation.
We investigate the first two components in \textit{computational performance analysis} and the last component in \textit{human effort analysis}.
All the following experiments are conducted using a single NVIDIA A6000 GPU, with results reported on the \textsc{Drebin} features and the MalRadar dataset.
}


\bfnoindent{\tcolor{Computational Performance Analysis}} \label{sec:computational_analysis}
\tcolor{Regarding the drift detection investment, we consider both the training and testing phases of the detector.
During training, our detector has similar computational overhead as CADE and remains more efficient than Transcend.
Specifically, the training computational complexity follows CADE's model, expressed as \( O(IB^2|\Psi|) \), where \( I \), \( B \), and \( |\Psi| \) represents the number of training iterations, batch size, and the number of model parameters. 
Although our detector is slightly larger than CADE's due to an additional dense layer for concept presence prediction, the small number of high-level concepts (e.g., 10 malicious behaviors) keeps the model sizes comparable, both estimated at $5.58$~MB; training time is slightly higher at $0.81$~s per epoch compared to $0.74$~s.
For Transcendent, evaluated in its rather computationally friendly setting with $10$-fold, has a model size of about $5.45$~MB but a longer training epoch time of $1.12$~s. 
During testing phase, where efficiency is important for real-time applications, \framework{} demonstrates superior performance by requiring only $0.57$~ms per sample without additional data. In contrast, CADE and Transcendent require $1.89$~ms and $5.75$~ms, respectively. Moreover, their dependence on training data increases memory usage.} 

\tcolor{In terms of operations for tuning the model, the concept-involved adaptor in \framework{} incurs the most computational cost. The retraining time is $2.86$~s per epoch, while the baseline adaptor takes $0.64$~s. 
Despite this, the time remains acceptable—fine-tuning requires much fewer iterations than training from scratch ($50$ epochs in our experiments)—and is unlikely to be a barrier to deployment~\cite{chen2023continuous}. The primary focus here is to achieve higher accuracy in the adapted model and reduce human effort, as will be explained subsequently.}




\bfnoindent{\tcolor{Human Effort Analysis}} \label{sec:human_effort_analysis}
The total human effort can be approximated by $n \times w$, where $n$ is the number of samples to be analyzed, and $w$ is the analysis workload per sample. 
\framework{} reduces human effort by decreasing 1)~$n$: utilizing an effective drift adaptor that requires significantly fewer budgets to achieve high accuracy~(\autoref{sec:eva_adaptation}), and 2)~$w$: providing an effective explainer that pinpoints key concepts faithfully~(\autoref{sec:eva_explanation}).
In the following, we further discuss the first aspect and support the second with \rcolor{a case study}. 

{Firstly, as shown in Figure~\ref{fig:reduced_budget}, we present the budget versus accuracy curves derived from previous adaptation performance results.
By selecting the desired classification accuracy threshold, we estimate the required labeling budget.
Typically, to achieve $90\%$ accuracy in the updated model, \framework{} requires only $19.54$ samples to be analyzed. In contrast, the baseline method requires $99.12$ samples, and even when enhanced with more time-consuming detectors, the best alternative still requires $83.67$ samples. This substantial reduction translates to $80.3\%$ savings in $n$.}

Secondly, to understand \framework{}'s explanation impact, we present a case study using drift explanations for a sample from \texttt{RuMMS}~(the largest family in the dataset), which mainly targets banking information through SMS interception~\cite{rumms16report}.
As illustrated in Figure~\ref{fig:drift_exp}, \framework{} detects \texttt{GhostCtrl} as the closest family in the training data.
While this family may serve as a reference, it focuses on remote control functionalities for data theft and manipulation~\cite{ghostctrl17report}, and our drift explainer identifies $3$ drifting behaviors out of the $10$ predefined concepts.
\rcolor{This explanation was manually verified to be useful in assigning the correct family label~(see \autoref{app:user_study} for a detailed discussion of its influence on analysts' workflows)}.
For instance, the identification of \textit{bank stealing} as a new behavior allowed analysts to effectively prioritize their investigation and find the primary malicious intent of the \texttt{RuMMS} malware. Furthermore, reviewing the reference pseudocode revealed that \textit{abusing phone calls} is also a significant drift: while \texttt{GhostCtrl} exhibited diverse and aggressive SMS/CALL functionalities, \texttt{RuMMS} specialized in SMS interception for data exfiltration. This specialization enabled analysts to understand the sample's more focused but narrower abusing behaviors.

%% file: empirical.tex
\begin{table}[tb]
    \centering
    \begin{tinytabularx}{\linewidth}{cCCCCCC}
    \toprule
    \textbf{Detector} & \textbf{2016} & \textbf{2017} & \textbf{2018} & \textbf{2019} & \textbf{2020} & \textbf{Avg.} \\
    \hline
    $\text{HCC}^\text{+}$ & 0.741 \newline \perc{4.75} & 0.718 \newline \perc{10.68} & 0.734 \newline \perc{21.35} & 0.765 \newline \perc{6.34} & 0.699 \newline \perc{6.80} & 0.731 \newline \perc{9.66} \\ 
    $\text{HCC}_{ce}^\text{+}$ & 0.745 \newline \perc{4.84} & 0.724 \newline \perc{11.52} & 0.719 \newline \perc{18.74} & 0.768 \newline \perc{6.81} & 0.698 \newline \perc{6.63} & 0.731 \newline \perc{9.43} \\ 
    $\text{HCC}_{hc}^\text{+}$ & 0.637 \newline \perc{9.61} & 0.700 \newline \perc{15.00} & 0.714 \newline \perc{20.59} & 0.713 \newline \perc{5.19} & 0.684 \newline \perc{4.08} & 0.690 \newline \perc{10.62} \\ \hline
    \framework{} & \textbf{0.754} \newline \perc{6.58} & \textbf{0.781} \newline \perc{20.31} & \textbf{0.866} \newline \perc{43.11} & \textbf{0.791} \newline \perc{9.98} & \textbf{0.763} \newline \perc{16.55} & \textbf{0.791} \newline \perc{18.57} \\
    \bottomrule
    \end{tinytabularx}
    \caption{The enhancement of intra-class drift detection with our detection framework.}
    \label{tab:intra-drift-detection}
    \vspace{-10pt}
\end{table}



To broaden our focus, we examine \framework{}'s effectiveness in intra-class drift.
\tcolor{In~\autoref{sec:motivation_lesson}, we evaluate drift detection methods from related work and demonstrate HCC's best overall performance on the dataset for binary malware detection.
Adopting the same experimental setup, we compare \framework{} against HCC in intra-class drift detection and adaptation.} 

\bfnoindent{Intra-class Drift Detection}
\framework{} achieves an $18.57\%$ AUC improvement on average compared to the HCC detector.
We further enhance the baseline with our insights about model sensitivity and the NCE metric, achieved by 
1)~the removal of the surrogate classifier $\mathbf{M}_s$ and its integration into the original classifier using the autoencoder structure; 2)~the augmentation of the previous cross-entropy based pseudo loss item with NCE~(\autoref{equ:nce_detector}).
As presented in~\autoref{tab:intra-drift-detection}, our modification achieves significant enhancements in HCC's detection performance, evident by both the integrated method and its two individual pseudo-loss detectors. Notably, there is an average performance increase from $9.43\%$ to $10.62\%$. The contrastive-based pseudo-loss detector, previously the least effective, now surpasses all prior baselines with the most substantial improvement. 
Nevertheless, our approach maintains an $8.21\%$ lead against the improved HCC method, largely due to our enriched latent space and data-autonomous design.

\bfnoindent{\tcolor{Intra-class Drift Adaptation}}
\tcolor{We integrate the HCC detector with the baseline adaptor to enable updates to external classifiers.
As presented in~\autoref{tab:hcc_dream_comparison}, \framework{} exceeds the HCC baseline in $76\%$ of cases, with more improvements observed at the higher labeling budget (i.e., 4.73\% at budget 100).
However, the gains are less pronounced compared to prior intra-class drift detection or our adaptor’s performance in inter-class scenarios, which can be attributed to two main reasons.
First, \framework{} detects more malware samples than the baseline ($51\%$ vs. $40\%$). 
Due to the highly imbalanced nature of binary malware detection ($10:1$ goodware to malware~\cite{pendlebury2019tesseract}), malware samples, despite being more prone to drift, represent a smaller portion of the retraining process.
Second, our adaptor’s design emphasizes embedding malware concepts to differentiate families. When adapting to a binary detection task, all benign samples are labeled with zero concept vectors.
Given the larger proportion of benign samples, this limits our semantic enrichment, resulting in less effective adaptation.}

\begin{table}[tb]
\centering
\begin{tinytabularx}{\linewidth}{cCCCCCC}
\toprule
\textbf{Budget} & \textbf{Adaptor} & \textbf{2016} & \textbf{2017} & \textbf{2018} & \textbf{2019} & \textbf{2020} \\ \hline
\multirow{2}{*}{10}  & HCC  & 0.9199 & 0.9151 & 0.8948 & 0.9070 & 0.8691 \\
    & \framework{} & \textbf{0.9206} & \textbf{0.9191} & \textbf{0.8959} & \textbf{0.9074} & \textbf{0.8708} \\ \hline
\multirow{2}{*}{20}  & HCC  & 0.9219 & 0.9204 & 0.8971 & 0.9088 & \textbf{0.8753} \\
    & \framework{} & \textbf{0.9228} & \textbf{0.9239} & \textbf{0.8982} & \textbf{0.9142} & 0.8733 \\ \hline
\multirow{2}{*}{30}  & HCC  & \textbf{0.9221} & 0.9221 & 0.8990 & 0.9090 & \textbf{0.8903} \\
    & \framework{} & 0.9196 & \textbf{0.9226} & \textbf{0.8998} & \textbf{0.9109} & 0.8870 \\ \hline
\multirow{2}{*}{40}  & HCC  & \textbf{0.9246} & 0.9217 & 0.8994 & \textbf{0.9128} & \textbf{0.8892} \\
    & \framework{} & 0.9208 & \textbf{0.9231} & \textbf{0.9001} & 0.9112 & 0.8857 \\ \hline
\multirow{2}{*}{100} & HCC  & 0.9241 & 0.9239 & 0.9017 & 0.9157 & 0.8890 \\
    & \framework{} & \textbf{0.9291} & \textbf{0.9304} & \textbf{0.9018} & \textbf{0.9176} & \textbf{0.8971} \\ \bottomrule
\end{tinytabularx}
\caption{\tcolor{\revOne{F1-score} comparison of HCC and \framework{} in intra-class drift adaptation across different labeling budget.}
\label{tab:hcc_dream_comparison}}
\vspace{-10pt}
\end{table}

%% file: discussion.tex
\section{Discussion}

\bfnoindent{Concepts and Annotators}
We define malicious Android behaviors as concepts to effectively \revOne{handle} inter-class drift. 
These concepts can be extended to cover benign functionality for intra-class drift and platform-specific behaviors beyond Android. 
\rcolor{For example, PE~(Windows) malware often exhibits behaviors such as process injection, registry persistence, and credential access~\cite{smith2020mind}. While some rootkit-like PE malware may involve deeper system-level integration, these behaviors still fall into structured categories (e.g., privilege escalation), similar to those in Android malware, so the concept space is not necessarily larger. 
To support generalization, we envision leveraging standardized taxonomies such as MITRE ATT\&CK~\cite{strom2018mitre}, which captures cross-platform adversarial techniques organized hierarchically, enabling both coarse- and fine-grained modeling of explicit concepts.}
To reduce labeling overhead, concept annotation can be supported by automated analysis tools~\cite{egele2008survey, lorenzoli2008automatic} or LLM-based methods~\cite{cai2023large, bansal2023large}, enhancing applicability across diverse malware triage settings.

\bfnoindent{Robustness to Attacks}
Autoencoders have shown promise in anomaly detection~\cite{hananomaly23owad} and have been extended to detect attacks.
For instance, CAE~\cite{razmi2023classification} applies a classifier in the autoencoder's latent space to identify poisoning samples, and MagNet~\cite{meng2017magnet} trains multiple autoencoders independently of the target classifier to detect adversarial attacks.
Interestingly, the autoencoder-based drift detector CADE~\cite{yang2021cade} is proven capable of detecting adversarial malware samples~\cite{astel2021}, and active learning is also promising in mitigating poisoning attacks~\cite{mcfadden2023rpal, lin2021active}.
Although attack defense is beyond our research scope, \tcolor{\framework{} combines autoencoder-based drift detection with active learning, incorporating both classifier and expert knowledge, potentially enhancing robustness against such attacks.}
We demonstrate \framework{}'s resilience in handling noisy labels in \autoref{app:open-world}, leaving a more in-depth analysis of attack robustness for future work.


%% file: related.tex
\section{Related Work} \label{sec:related}

\bfnoindent{OOD Detection}
The majority of existing OOD detection methods in machine learning community rely on auxiliary OOD dataset~\cite{hendrycks2016baseline,chen2020robust,liang2017enhancing,masana2018metric}. 
For instance, Chen et al.~\cite{chen2020robust} uses an auxiliary dataset like the 80 Million Tiny Images. 
However, acquiring such large-scale and comprehensive auxiliary OOD datasets can be particularly challenging in the malware domain.
Similar to existing work in malware domain, we are not dependent on auxiliary OOD dataset to offer greater practicality.
Moreover, \framework{} operates independently of any training data during the drift detection phase, enhancing its applicability in real-world scenarios.

\bfnoindent{Labelless Drift Adaptation} 
Besides active learning, there are also drift adaptation strategies without labels~\cite{kan2021investigating}. 
For example, APIGraph~\cite{zhang2020enhancing} and AMDASE~\cite{yang2024novel} use semantically-equivalent API usages to mitigate classifier aging~\cite{xu2019droidevolver}. 
However, they only \revOne{apply} to classifiers with specific types of features.
DroidEvolver~\cite{xu2019droidevolver, kan2021investigating} employs pseudo-labels and is a promising solution to address labeling capacity.
Nevertheless, this method can easily lead to negative feedback loops and self-poisoning~\cite{barbero2022transcending}.
Our work is based on active learning which involves human and effectively minimizes the labeling budget.
Nevertheless, these works can be complementary to us to further enhance the robustness in drift adaptation.

\bfnoindent{Explainable Security Applications}
Recent research has focused on offering post-hoc explanations to security applications. For example, in malware detection~\cite{anderson2018ember,arp2014drebin} tasks, FINER~\cite{he2023finer} produces function-level explanations to facilitate code analysis. 
In malware mutation~\cite{lakshminarayanan2017simple,chen2015finding} applications, AIRS~\cite{yu2023airs} explains deep reinforcement learning models in security by offering step-level explanations. 
On a different basis, \framework{} addresses the explainability problem in a drift adaptation setting, where the intrinsic behavioral explanations can propagate expert revisions to update the classifier.

\bfnoindent{Explanatory Interactive Learning}
Recent advancements in machine learning have combined explainable AI with active learning, leading to progress in explanatory interactive learning~\cite{stammer2021right,selvaraju2019taking,ross2017right}. 
Inspired by this, our approach incorporates human feedback on both labels and explanations, but with distinct objectives and settings. 
Unlike these works that focus on image domain and feature-level explanation annotation, our method is tailored for malware analysis and generates high-level explanations. 
Furthermore, while they use external explainers to guide classifiers~\cite{stammer2021right}, our approach integrates explanations directly within the drift detector, enhancing the efficiency of our drift adaptation method.

%% file: conclusion.tex
To deploy deep learning-based malware classifiers in dynamic and hostile environments, our work addresses a crucial aspect of combating concept drift.
The proposed \framework{} system emerges as an innovative and effective solution. 
Behavioral explanations are integrated into a contrastive autoencoder, connecting the classifier to achieve model-sensitive detection and explanatory adaptation.
The effectiveness of \framework{} against evolving threats is demonstrated through extensive evaluation and marks a notable advancement over existing methods. 
We make the system open-source~\footnote{\url{https://github.com/E0HYL/DREAM-drift-adapt}} and hope that it can inspire future research to explore concept drift in broader security contexts.

%% file: appendix.tex
\begin{figure*}[tb]
\centering
    \subfloat[Transcendent, \textsc{Drebin}]{%
        \includegraphics[width=0.24\linewidth]{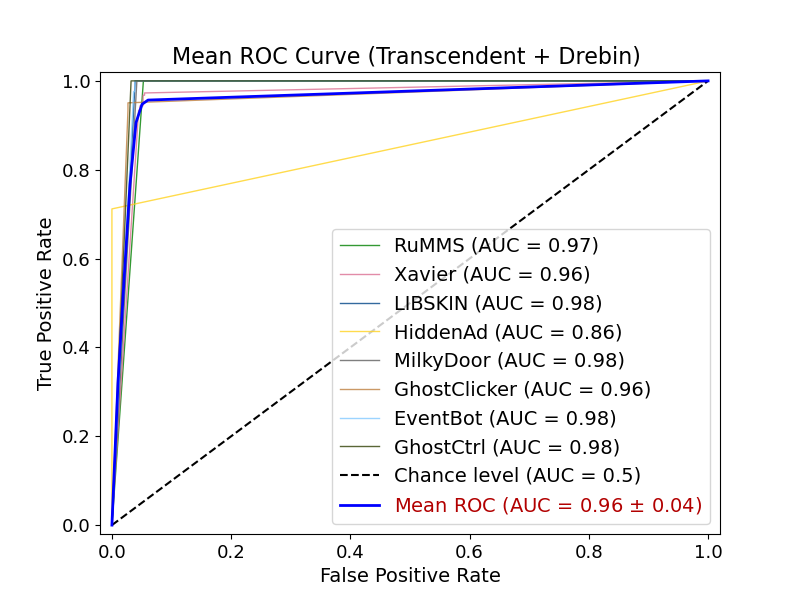}%
        \label{fig:malradar_a}}%
    \hfill
    \subfloat[CADE, \textsc{Drebin}]{%
        \includegraphics[width=0.24\linewidth]{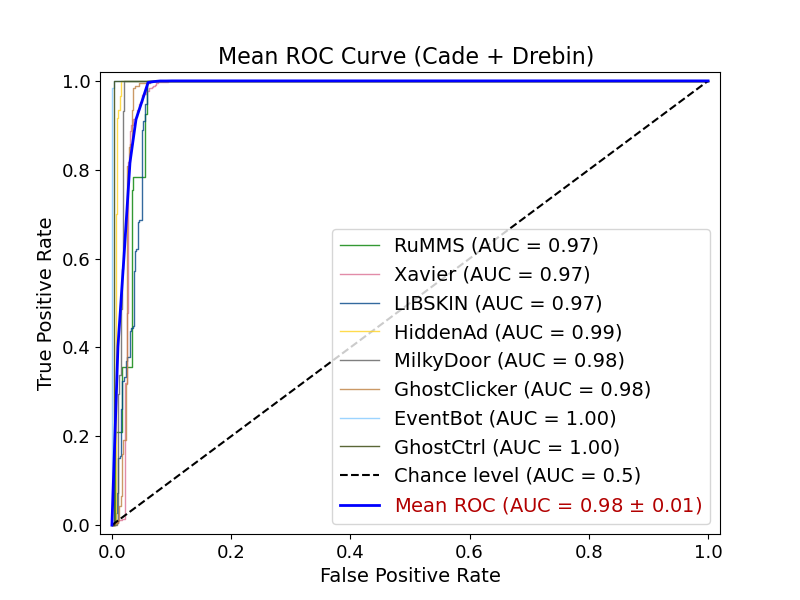}%
        \label{fig:malradar_b}}%
    \hfill
    \subfloat[Probability, \textsc{Drebin}]{%
        \includegraphics[width=0.24\linewidth]{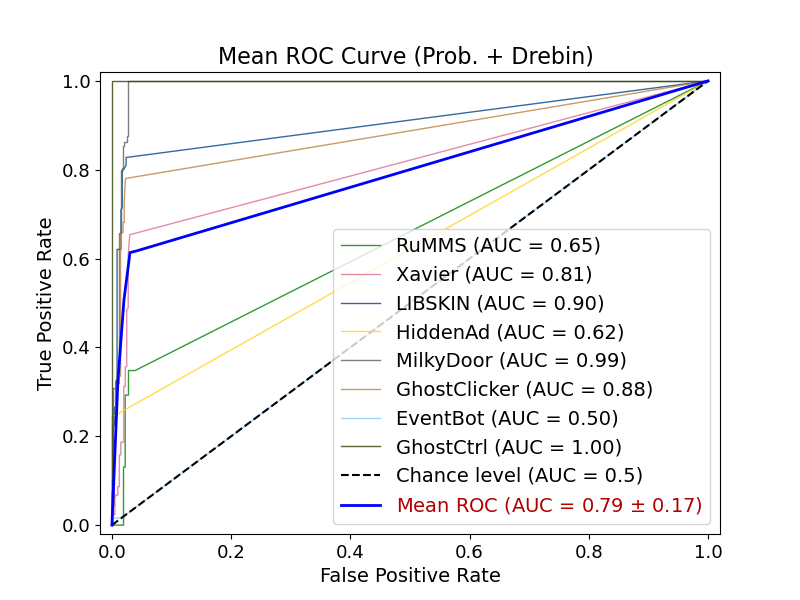}%
        \label{fig:malradar_c}}%
    \hfill
    \subfloat[(Ours) \framework{}, \textsc{Drebin}]{%
        \includegraphics[width=0.24\linewidth]{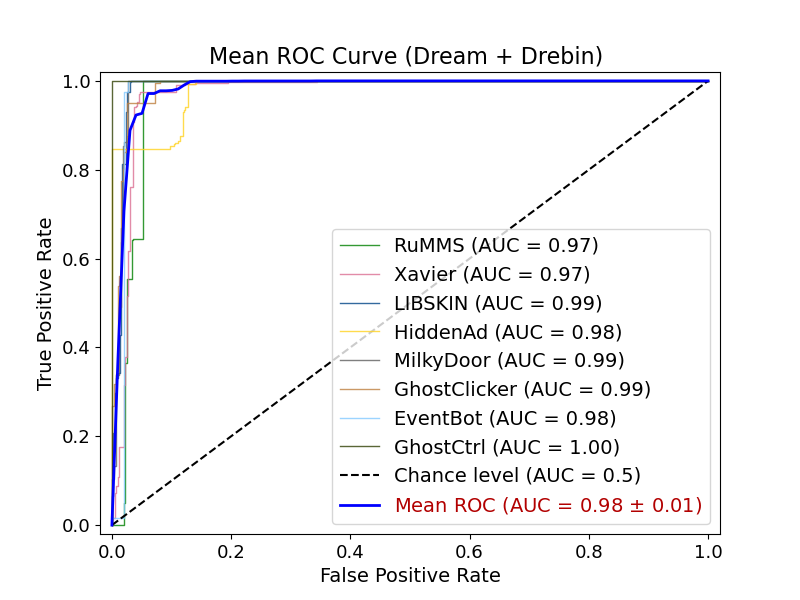}%
        \label{fig:malradar_d}}
    \vspace{-1em}
    
    \subfloat[Transcendent, \textsc{Mamadroid}]{%
        \includegraphics[width=0.24\linewidth]{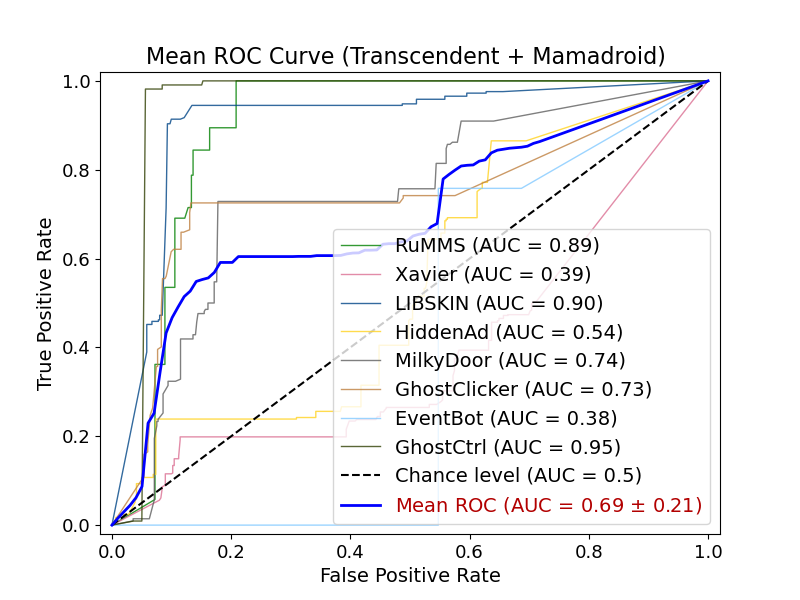}%
        \label{fig:malradar_e}}%
    \hfill
    \subfloat[CADE, \textsc{Mamadroid}]{%
        \includegraphics[width=0.24\linewidth]{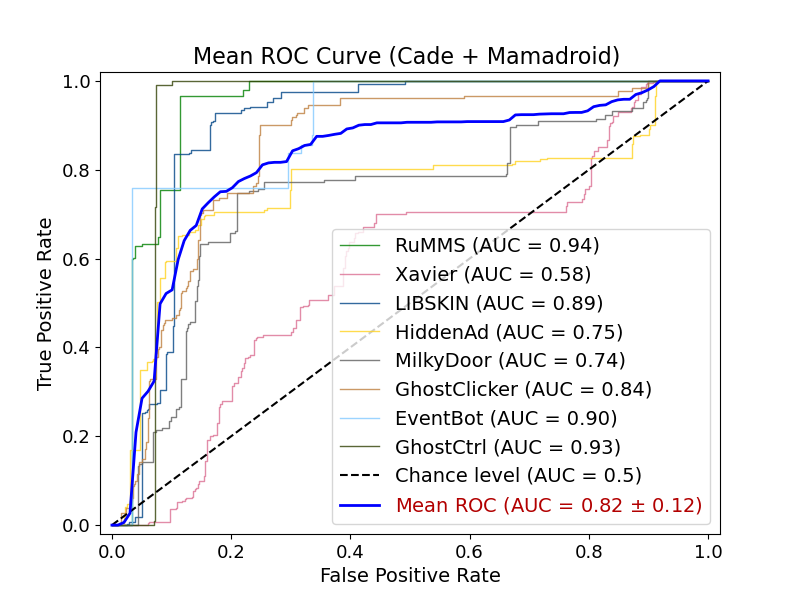}%
        \label{fig:malradar_f}}%
    \hfill
    \subfloat[Probability, \textsc{Mamadroid}]{%
        \includegraphics[width=0.24\linewidth]{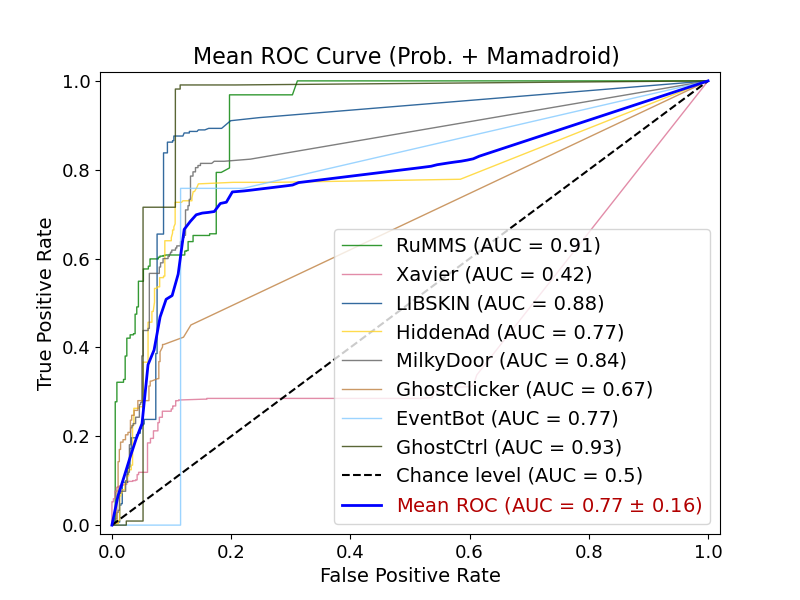}%
        \label{fig:malradar_g}}%
    \hfill
    \subfloat[(Ours) \framework{}, \textsc{Mamadroid}]{%
        \includegraphics[width=0.24\linewidth]{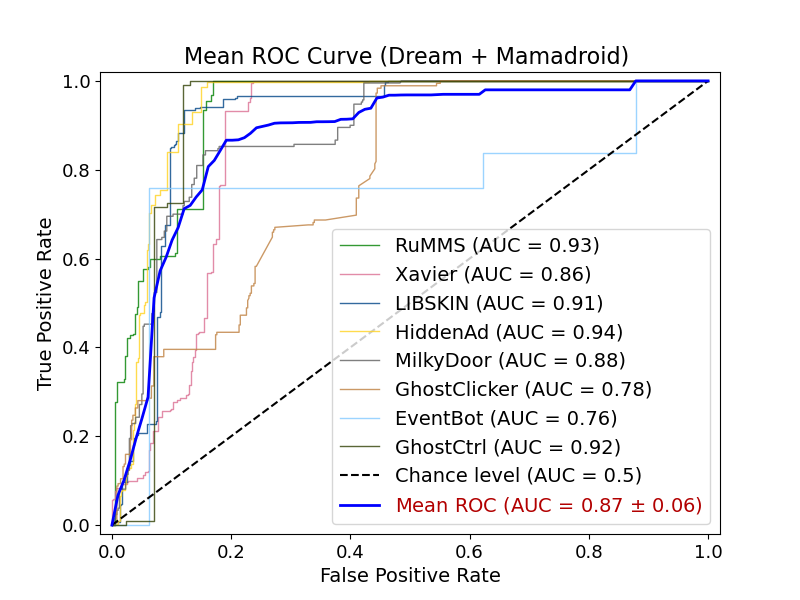}%
        \label{fig:malradar_h}}
    \vspace{-1em}
    
    \subfloat[Transcendent, \textsc{Damd}]{%
        \includegraphics[width=0.24\linewidth]{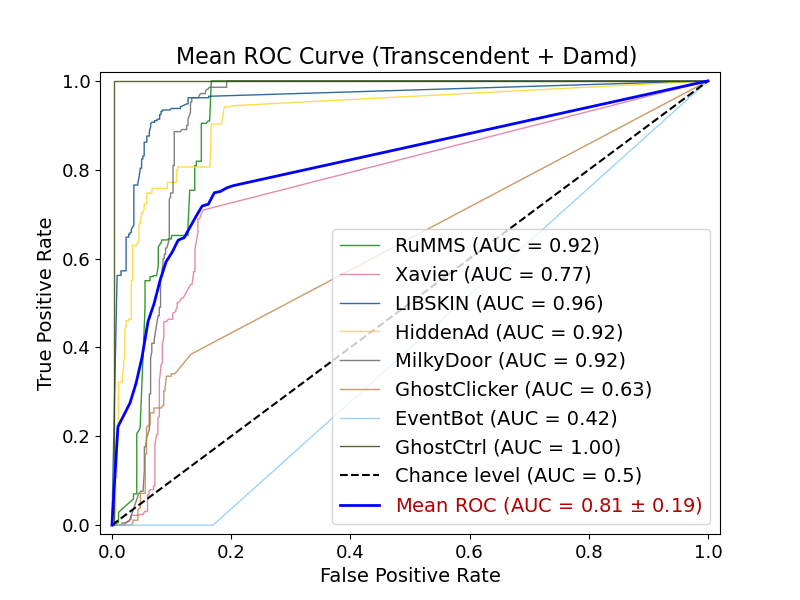}%
        \label{fig:malradar_i}}%
    \hfill
    \subfloat[CADE, \textsc{Damd}]{%
        \includegraphics[width=0.24\linewidth]{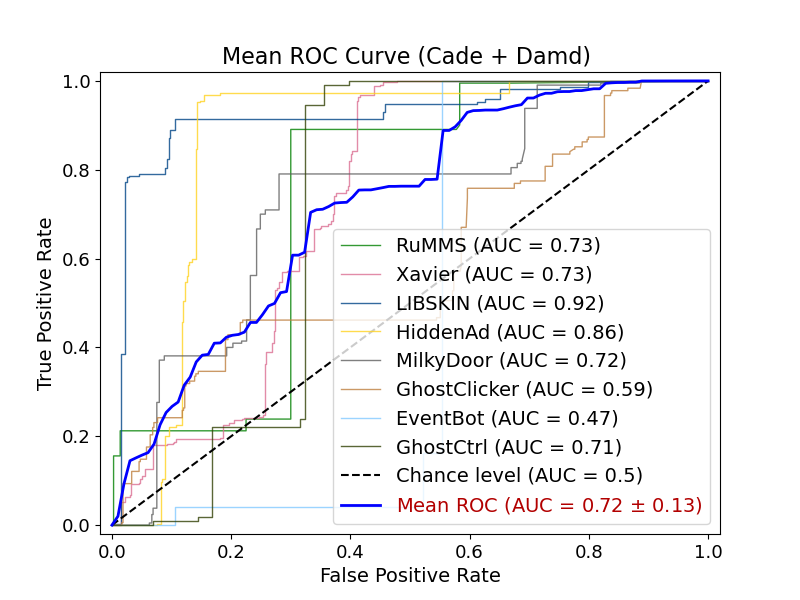}%
        \label{fig:malradar_j}}%
    \hfill
    \subfloat[Probability, \textsc{Damd}]{%
        \includegraphics[width=0.24\linewidth]{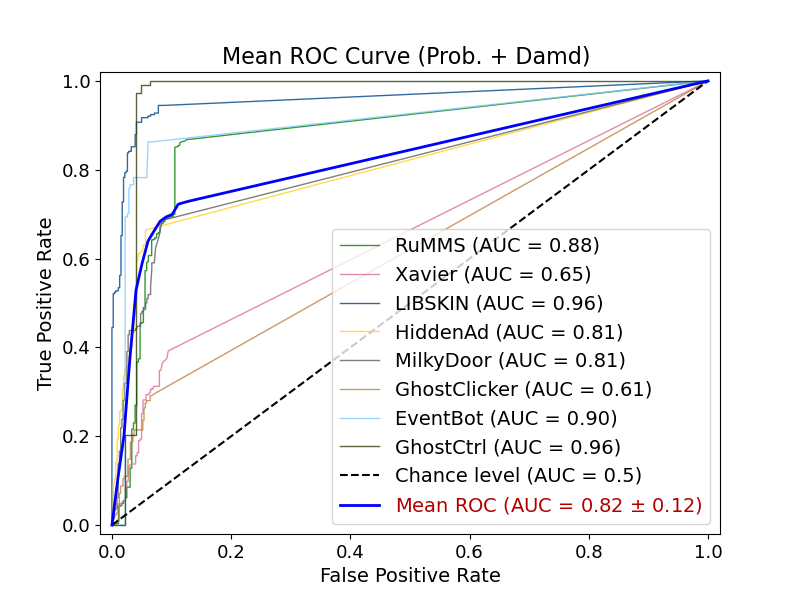}%
        \label{fig:malradar_k}}%
    \hfill
    \subfloat[(Ours) \framework{}, \textsc{Damd}]{%
        \includegraphics[width=0.24\linewidth]{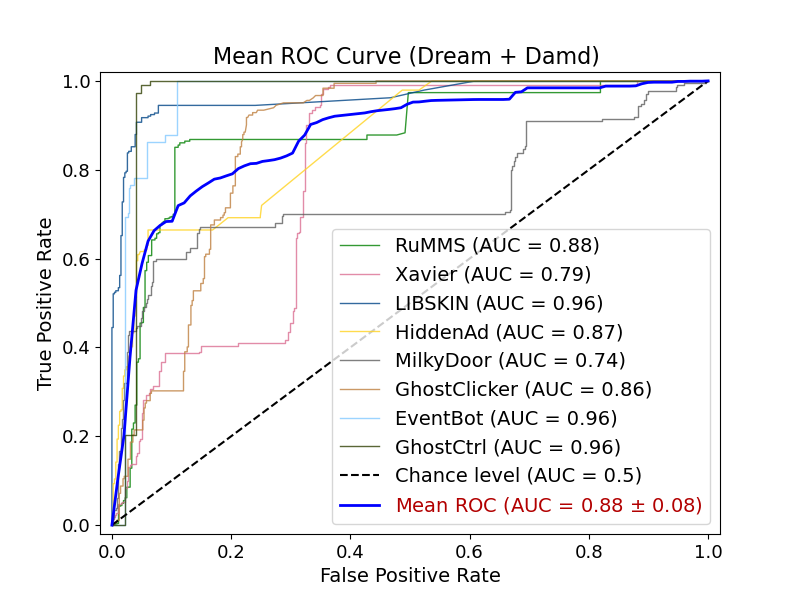}%
        \label{fig:malradar_l}}
    \vspace{-.8em}
    \caption{Evaluation of inter-class drift detection on Malradar dataset with three feature spaces. The first three columns are Transcendent, CADE, Probability, respectively, and our method is on the last column.} %
    \label{fig:malradar_multi-detection-basic}
\end{figure*}

\section{Malware Dataset} \label{app:dataset}

As shown in \autoref{tab:drebin_malradar}, we use two datasets for malware family classification.
Specifically, we select 8 families for each of the Drebin~\cite{yang2021cade} and the MalRadar~\cite{wang2022malradar} dataset, where the family selection \revOne{adheres} to the same criteria used in the CADE paper.  
The Drebin dataset spans the years from 2010 to 2012, offering insights into malware characteristics from the \revOne{early} period.
To capture more recent trends in the evolving malware landscape, the Malradar dataset \cite{wang2022malradar} covers the period from 2015 to 2021. 

We use the malicious behaviors defined in the Malradar paper.
To ensure consistency in behavioral analysis, we augment the Drebin dataset, which originally lacks behavior labels, with these same behaviors. 
This extension involves extrapolating the labels based on Malradar's definitions, supplemented by expert analysis and validation through GPT~\cite{pa2023attacker}. 
Note that the concept labels were assigned with minimal additional effort, and it is uniform within each family. 
In cases where certain behaviors in the Malradar dataset are specific to particular variants, we accommodate this by setting the relevant valid label mask $m_e$ to zero. 
Despite the simplicity, this approach is in line with an active learning setting that typically utilizes limited human effort.

For the Drebin dataset, the time-split is performed across all remaining families, as detailed in the CADE paper. 
For the Malradar dataset, its extensive time range and emergence of new families in recent years necessitate a tailored splitting approach. The previous split would classify newer families, such as \texttt{EventBot} which emerged in 2020, entirely into the testing set, potentially introducing unintended drift. To avoid this, we adjust the strategy by performing the split on a per-family basis within this dataset.

\section{Autoencoder Architecture} \label{app:autoencoder}
For the \textsc{Drebin} feature, the autoencoder used is a form of Tabular Autoencoder consisting of two dense layers in both the encoder and decoder components. The architecture features a hidden dimension size of $512$ and an encoding dimension of $32$.
In the case of the \textsc{Mamadroid} feature, the autoencoder is similarly structured as a Tabular Autoencoder, but with a larger hidden dimension of $2048$ and an encoding dimension of $128$. This expanded architecture accommodates the more complex nature of the Mamadroid feature set.
For the \textsc{Damd} feature, the autoencoder is a specialized convolutional Text Autoencoder. 
This autoencoder works in conjunction with an embedding layer to facilitate reconstruction on numerical data. 
The encoder incorporates a convolutional layer with a kernel size of $3$ and $64$ filters, followed by a global max pooling layer. 
The decoder comprises a dense layer and a convolutional transpose layer that transform the data back into the embedding dimension, maintaining the same kernel size. 
This configuration is tailored to effectively handle the sequential nature of this feature set.

\input{revision/open_world_testing}

\input{revision/computation_exp_appendix}

\input{appendix/id_explain}

\section{Baselines from Existing Work} \label{sec:app_existing}
Advancements in drift detection and adaptation techniques are crucial for combating malware drifts effectively. 
Focusing on these aspects, three notable studies have contributed significantly to the field. 
As outlined in \autoref{tab:related}, we discuss their methodologies and comparative research focuses below, leaving technical details of their detectors in~\autoref{sec:formalized_detectors}.

\bfnoindent{Transcendent} 
Transcendent~\cite{barbero2022transcending} innovates the conformal prediction-based drift detection by introducing novel conformal evaluators that refine the calibration process.
Compared to its predecessor~\cite{jordaney2017transcend}, this refinement allows for a much more efficient calculation of p-values, concurrently enhancing drift detection accuracy.
This method does not introduce a new detection model and can be generally applied to different types of drift and classifier architectures.
However, it relies on statistical analysis with frequent retraining of the classifier~(which can be time-consuming for complex classifiers) and its drift adaptation is conservative as it simply rejects drift samples.

\bfnoindent{CADE} 
Contrastive learning is introduced to the nonconformity scoring based drift detection by CADE~\cite{yang2021cade}.
It trains an unsupervised autoencoder to create a latent space for measuring distances, and the nonconformity for a test sample is the minimum distance to the multi-class centroids of training data.
This work also pioneers in drift explanation by connecting drift detection decisions to important features, but it does not address how explanations integrate into the updating process of the classifier~\cite{Roberts18Explanatory, teso2019explanatory}.
For adaptation, it simply applies retraining and is limited to intra-class scenarios. 

\bfnoindent{HCC} 
Hierarchical Contrastive Classifier (HCC) presents a novel malware classifier by implementing a dual subnetwork architecture~\cite{chen2023continuous}. 
The first subnetwork leverages contrastive learning to generate embeddings, which are then utilized by the second for malware detection. 
HCC integrates active learning and improves CADE in intra-class scenarios.
Specifically, it customizes intra-class by infusing a hierarchical design in the contrastive loss~\cite{zhang2022use} and defining a pseudo loss to capture model uncertainty with training data.
Nevertheless, a notable feature of the HCC detector is its inherent design, which can pose challenges in adapting off-the-shelf classifiers.

\input{revision/adaptation_detail_appendix}

\input{overview}

%% file: revision/open_world_testing.tex
\section{Open-world Testing} \label{app:open-world}

In this section, we present two case studies to evaluate the generalization and robustness of our approach in open-world scenarios. Specifically, we explore the detector's ability to handle various unseen test families and the adaptor's resilience to imperfect human feedback during active learning.

\begin{table}[tb]
    \begin{tinytabularx}{.99\linewidth}{*{6}Cc}
    \toprule
           & \textbf{f0+}     & \textbf{f1+}     & \textbf{f2+}      & \textbf{f3+}     & \textbf{f4+}     & \textbf{Avg. $\pm$ Std.}  \\ \hline
    \textbf{Prob.}  & -3.4\% & 23.1\% & -12.5\% & 8.6\%  & -9.9\% & 1.2\% $\pm$ 13.2\% \\
    \textbf{Trans.} & -3.3\% & 14.3\% & -9.4\%  & 4.3\%  & 1.1\%  & 1.4\% $\pm$ 7.9\% \\
    \textbf{CADE}   & -1.4\% & 13.7\% & -14.1\% & -4.7\% & 12.5\% & 1.2\% $\pm$ 10.6\% \\
    \textbf{\framework{}}  & 0.0\%  & 11.4\% & -5.2\%  & 5.7\%  & -4.1\% & 1.6\% $\pm$ 6.2\%\\
    \bottomrule
    \end{tinytabularx}
    \caption{\revTwo{Impact of including small families in the test set: percentage change in drift detection AUC. The columns labeled f0+ to f4+ indicate results where, in addition to the held-out family used as drift, previously excluded small families in MalRadar are also included in the test set.}}
    \label{tab:small_family_detection}
    \vspace{-5pt}
\end{table}

\bfnoindent{\revTwo{Small Families}}
\revTwo{To evaluate the drift detector's ability to generalize beyond large families, we assess its performance when previously excluded small families from the MalRadar dataset are added to the test set. 
In our earlier setup, each classifier was trained by holding out one of the largest families, which served as the drift source during testing.
Here, we extend this setting by introducing $1,821$ additional samples from $172$ small families into the test set; these samples are also treated as drift.
This experiment focuses on the top five largest families and is conducted using the DAMD feature representation. 
\autoref{tab:small_family_detection} reports the percentage change in drift detection AUC resulting from the inclusion of small-family samples.
Interestingly, the inclusion does not necessarily degrade performance. 
On average, all methods maintain stable detection AUC compared to the original setup~(see \autoref{fig:malradar_multi-detection-basic}), and in some cases, even achieve slightly higher performance. 
This stability arises because, although small families differ from the held-out drift family, their samples remain outside the training distribution and are still detectable as drift.
Notably, our method shows both the highest average AUC and the most stable performance across models (i.e., lowest standard deviation), demonstrating its generalization in a more open-world testing scenario.
}

\bfnoindent{Labeling Noise}
We conduct a case study to compare the resilience of our method and the benchmark against labeling \revOne{noise}~\cite{pirch2021tagvet}, where we use the model trained on the \textsc{Drebin} feature of the Drebin dataset and a labeling budget of 50.
\revOne{Noise is introduced by randomly selecting a subset of samples and reassigning their family labels to the nearest families.}
As shown in \autoref{tab:noise_adaptaion}, DREAM consistently achieves higher and more stable F1 and accuracy scores compared to the benchmark. 
Moreover, when the noise level increases, such as to 6\%, the accuracy score improvement can reach 28.4\%. This could be attributed to the inclusion of an explanation revision step during model updates, making our method more robust and resilient to noise.
\revOne{To examine where DREAM’s performance begins to degrade, we extend the noise ratio and observe that DREAM remains robust up to 30–40\% noise, maintaining F1-scores above 0.8. Beyond this point, a noticeable drop occurs (e.g., 0.711 at 50\% noise), indicating the method’s practical tolerance to moderate annotation errors.}

\begin{table}[tb]
    \centering
    \begin{tinytabularx}{.99\linewidth}{*{7}C}
    \toprule
    \multirow{2}{*}{\begin{tabular}{@{}c@{}} \textbf{Noise} \\ \textbf{Ratio} \end{tabular}} & \multicolumn{3}{c}{\textbf{F1-score}} & \multicolumn{3}{c}{\textbf{Accuracy}} \\ \cmidrule(lr){2-4} \cmidrule(lr){5-7}
     & \revOne{\textbf{Bench.}} & \textbf{Ours} & \textbf{Imp.} & \revOne{\textbf{Bench.}} & \textbf{Ours} & \textbf{Imp.} \\ \hline
    0\% & 0.894 & \textbf{0.966} & \cellcolor[HTML]{FFFFFF}8.0\% & 0.863 & \textbf{0.956} & \cellcolor[HTML]{FFFFFF}10.7\% \\
    2\% & 0.862 & \textbf{0.950} & \cellcolor[HTML]{FFFFFF}10.2\% & 0.800 & \textbf{0.916} & \cellcolor[HTML]{FFFFFF}14.5\% \\
    6\% & 0.801 & \textbf{0.967} & \cellcolor[HTML]{FFFFFF}20.7\% & 0.760 & \textbf{0.976} & \cellcolor[HTML]{FFFFFF}28.4\% \\
    10\% & 0.822 & \textbf{0.945} & \cellcolor[HTML]{FFFFFF}14.9\% & 0.772 & \textbf{0.959} & \cellcolor[HTML]{FFFFFF}24.3\% \\
    \bottomrule
    \end{tinytabularx}
    \caption{Impact of labeling noise during active learning: comparison of drift adaptation results with the benchmark under different noise ratios.}
    \label{tab:noise_adaptaion}
    \vspace{-5pt}
\end{table}

%% file: revision/computation_exp_appendix.tex
\section{Insights from Drift Explanation} \label{app:user_study}

\begin{figure}[tb]
    \centering
    \includegraphics[width=\linewidth]{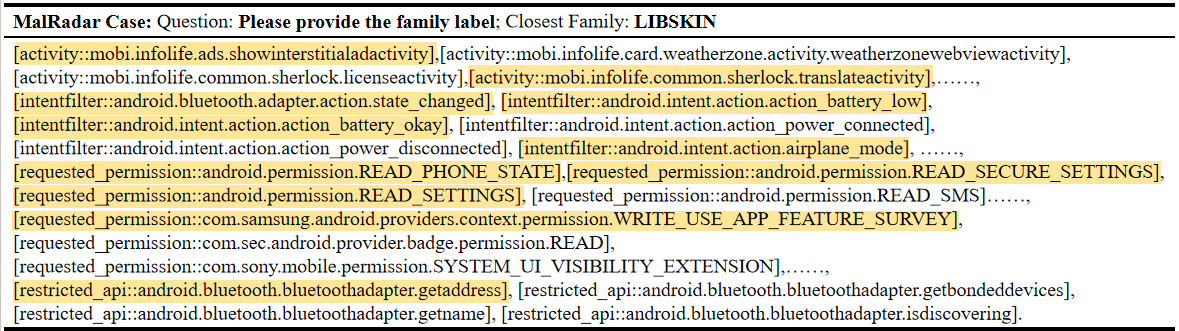}
    \caption{\tcolor{CADE's drift explanation example~(truncated version; original length $1329$) for a sample from the \texttt{RuMMS} family, highlighting features contributing to drift compared to its `Closest Family' output. \textsc{Drebin} feature space is sparse and only $28$ features differ by diff, but $29$ are highlighted and only $7$ correctly match the actual differences.}}
    \label{fig:cade_drift_exp}
\end{figure}

Unlike traditional explanations for ID data, drift explanations highlight differences between new samples and existing training data, posing unique challenges for effective presentation and user studies~\cite{nadeem2023sok}.
Previous approaches fail to convey the reference-based properties, simply visualizing important features on drift samples~\cite{yang2021cade}~(\autoref{fig:cade_drift_exp}). 
We address the shortcoming by connecting drift explanations with the original behaviors of the closest family, with an example presented in Figure~\autoref{fig:drift_exp}\revOne{.~\footnote{SHA256: 7ae4f72eeec59d03a20fda4a04f8098875df20af6d9f7e3625d1800bbe169ec3}}
\rcolor{In this case study, the explanation has the potential to guide malware analysis in three major steps:}
\begin{itemize}
    \item Behavior Identification: analysts can combine behaviors from the closest family with the identified drift behaviors, focusing on four specific behaviors out of ten categories. The non-drift behavior \emph{remote control} prompts further examination for similar malicious activities in the new sample.
    \item Prioritization of New Behaviors: the identification of the new behavior, \emph{bank stealing}, allows analysts to prioritize their investigation effectively, recognizing it as a particularly dangerous activity. Around these functions, an initial understanding of two other malicious behaviors can also be developed.
    \item Drift Verification of Uncertain Behaviors: to verify the behavior \emph{abusing SMS/CALL} as drifting, analysts can review the reference pseudocode and find that, unlike \texttt{GhostCtrl}'s broad SMS functionalities, the new sample specifically focuses on SMS interception for data exfiltration. Similarly, \emph{privacy stealing} is narrower compared to \texttt{GhostCtrl}'s broader device profiling.
\end{itemize}

\tcolor{Overall, \framework{}'s explanations may help analysts narrow their focus with a reasonable reference and drift behaviors, enabling them to efficiently identify similarities and significant behavioral differences, ultimately reducing their labeling workload.}

%% file: appendix/id_explain.tex
\section{Concept-based ID Explanation}

The drift explainer in our system can explain OOD data, identifying concepts which contributes most to the drifting. 
Besides drift explanation, \framework{} leverages an autoencoder intermediary to generate concept-based explanations for in-depth analysis of ID data. 
The challenge in offering high-level explanations, beyond mere feature attribution, persists even in non-drifting scenarios. Current methodologies tend to abstract explanations from features~\cite{ribeiro2016should, he2023finer}, but they cannot readily translate into behavioral insights.


To establish a baseline for concept-based explanations, we frame it as a series of binary classification tasks, each corresponding to a different explicit concept. Specifically, we adapt the output layer of the classifier to yield sigmoid probabilities and conduct fine-tuning over 100 retraining epochs. This ensures that each classifier achieves an accuracy of over 0.99 during training. We then select in-distribution data from the test dataset—specifically, data that corresponds to classes previously encountered during training—and assess the concept classification accuracy across all concepts.

We conduct the experiments with the Drebin dataset, and as in~\autoref{tab:id_concept_exp}, the results reveal that our method achieves near-perfect accuracy on the non-drifting test dataset, approaching $100\%$. This represents a significant $93.9\%$ improvement over the baseline across all models, indicating strong stability on in-distribution data.
However, we observed a stark decrease in concept accuracy on out-of-distribution data, with average accuracies around $66\%$ for both the baseline and our method. 
Interestingly, the baseline shows a slight improvement in this context, possibly due to its conservative nature, leading to more frequent negative outputs.
These observations emphasize the necessity of updating both the detector and classifier during updating, a key aspect of our drift adaptor's design.



\begin{table}[tb]
    \centering
    \begin{smalltabularx}{.8\linewidth}{cCCc}
    \toprule
    \textbf{Hold-out} & \textbf{Baseline} & \textbf{Ours} & \textbf{Imp.} \\ \midrule
    FakeInstaller & 0.472 & 0.999 & 111.8\% \\
    DroidKungFu & 0.496 & 1.000 & 101.8\% \\
    Plankton & 0.516 & 1.000 & 93.9\% \\
    GingerMaster & 0.526 & 1.000 & 90.2\% \\
    BaseBridge & 0.527 & 1.000 & 89.8\% \\
    Iconosys & 0.532 & 1.000 & 87.9\% \\
    Kmin & 0.532 & 1.000 & 87.8\% \\
    FakeDoc & 0.532 & 1.000 & 87.8\% \\
    \bottomrule
    \end{smalltabularx}
    \caption{Concept classification accuracy for ID test data.}
    \label{tab:id_concept_exp}
    \vspace{-5pt}
\end{table}

\begin{table}[tb]
    \centering
    \begin{tinytabularx}{.68\linewidth}{cCC}
    \toprule
        \textbf{Explainer} & \textbf{CBP} & \textbf{DRR} \\ \hline
        Random & 0.017 ± 0.031 & 0.023 ± 0.009 \\
        Dri-IG & 0.228 ± 0.103 & 0.355 ± 0.337 \\
        CADE & 0.173 ± 0.112 & \textbf{\text{0.974 ± 0.009}} \\ 
        \framework{} & \textbf{\text{0.331 ± 0.197}} & \textbf{\text{0.974 ± 0.009}} \\
    \bottomrule
    \end{tinytabularx}
    \caption{Evaluation of feature-level explanations.}
    \label{tab:feature_explainer_eva}
    \vspace{-5pt}
\end{table}

%% file: revision/adaptation_detail_appendix.tex

\begin{table}[tb]
    \centering
    \resizebox{.98\columnwidth}{!}{%
    \begin{tinytabularx}{\linewidth}{p{0.55cm}c|CCCC|CCCC}
    \toprule
     \multirow{2}{*}{\textbf{Feature}} & \multirow{2}{*}{\textbf{\#N}} & \multicolumn{4}{c|}{\textbf{Drebin}} & \multicolumn{4}{c}{\textbf{MalRadar}} \\ \cmidrule(lr){3-6} \cmidrule(lr){7-10}
    && \footnotesize{Trans.} & \footnotesize{CADE} & \footnotesize{Prob.} & \footnotesize{CRD} & \footnotesize{Trans.} & \footnotesize{CADE} & \footnotesize{Prob.} & \footnotesize{CRD} \\ \hline
    
     \multirow{5}{*}{\textsc{Drebin}} & 10 & 2.0 & 9.3 & \textbf{9.9} & \cellcolor[HTML]{EFEFEF}9.6 & \cellcolor[HTML]{EFEFEF}5.0 & 6.8 & 7.9 & \textbf{8.4} \\ 
     
     & 20 & \cellcolor[HTML]{EFEFEF}5.8 & 19.0 & \textbf{19.4} & 19.0 & \cellcolor[HTML]{EFEFEF}11.8 & 15.9 & 16.4 & \textbf{17.9} \\ 
     
     & 30 & \cellcolor[HTML]{EFEFEF}9.5 & 28.9 & \textbf{29.4} & 29.0 & \cellcolor[HTML]{EFEFEF}21.8 & 25.8 & 24.9 & \textbf{27.6} \\ 
     
     & 40 & \cellcolor[HTML]{EFEFEF}13.8 & 38.8 & \textbf{39.4} & 39.0 & \cellcolor[HTML]{EFEFEF}31.8 & 35.6 & 33.5 & \textbf{37.5} \\ 
     
    & 100 & 44.4 & \cellcolor[HTML]{EFEFEF}\textbf{98.5} & 98.3 & 97.9 & 91.8 & 94.6 & 80.3 & \cellcolor[HTML]{EFEFEF}\textbf{96.3} \\ \hline
    
     \multirow{5}{*}{\makecell{\textsc{Mama-} \\ \textsc{droid}}}& 10 & \textbf{8.6} & \cellcolor[HTML]{EFEFEF}5.6 & 6.8 & 5.6 & 4.4 & 0.5 & \cellcolor[HTML]{EFEFEF}4.8 & \textbf{4.9} \\ 
     
     & 20 & \textbf{17.4} & \cellcolor[HTML]{EFEFEF}9.3 & 14.8 & 12.5 & 8.9 & 3.8 & \textbf{11.9} & \cellcolor[HTML]{EFEFEF}10.0 \\ 
     
     & 30 & \textbf{24.0} & 12.9 & 23.0 & \cellcolor[HTML]{EFEFEF}20.9 & 14.1 & 10.0 & \cellcolor[HTML]{EFEFEF}\textbf{18.8} & 15.3 \\ 
     
     & 40 & \cellcolor[HTML]{EFEFEF}30.3 & 17.9 & \textbf{31.1} & 28.5 & 20.9 & 17.6 & \cellcolor[HTML]{EFEFEF}\textbf{25.8} & 21.9 \\ 
     
     & 100 & \cellcolor[HTML]{EFEFEF}70.4 & 46.5 & 78.0 & \textbf{79.5} & 62.4 & 68.4 & \textbf{71.9} & \cellcolor[HTML]{EFEFEF}68.3 \\  \hline
    
     \multirow{5}{*}{\textsc{Damd}}& 10 & \cellcolor[HTML]{EFEFEF}7.0 & 3.8 & 7.8 & \textbf{8.1} & 4.1 & 2.1 & \cellcolor[HTML]{EFEFEF}5.4 & \textbf{5.6} \\ 
     
     & 20 & \cellcolor[HTML]{EFEFEF}13.9 & 6.1 & 14.9 & \textbf{15.1} & 10.4 & 7.1 & \textbf{14.8} & \cellcolor[HTML]{EFEFEF}14.6 \\ 
     
     & 30 & \cellcolor[HTML]{EFEFEF}20.8 & 10.3 & \textbf{21.1} & 20.5 & 17.5 & 11.6 & \cellcolor[HTML]{EFEFEF}\textbf{24.1} & \textbf{24.1} \\ 
     & 40 & \cellcolor[HTML]{EFEFEF}27.8 & 13.4 & \textbf{28.5} & 26.1 & 24.6 & 16.9 & \cellcolor[HTML]{EFEFEF}\textbf{32.2} & \cellcolor[HTML]{EFEFEF}31.9 \\ 
     & 100 & \cellcolor[HTML]{EFEFEF}68.6 & 27.0 & 68.4 & \textbf{74.1} & 71.5 & 56.5 & \textbf{84.5} & \cellcolor[HTML]{EFEFEF}84.0 \\
    
    \bottomrule
    \end{tinytabularx}
    }
    \caption{\tcolor{Number of test samples correctly identified and used for adaptation. Numbers in bold indicate the highest counts. Cells in grey indicate that the detector achieves the highest F1-score paired with the baseline adaptor.}}
    \label{tab:adaptation details}
    \vspace{-5pt}
\end{table}

%% file: overview.tex
\section{Formalization} \label{sec:formalization}

\bfnoindent{Notations}
We define an input instance \( \mathbf{x} \) as an element of the feature space \( \mathcal{X} \subseteq \mathbb{R}^{p \times q} \), where \( p \) and \( q \) represent dimensions pertinent to the attributes of the data.
The label for any instance \( \mathbf{x} \) is represented by \( y \), where \( y \) belongs to the label space \( \mathcal{Y} \). 
The space \( \mathcal{Y} \) can be binary, for instance, \( \{0, 1\} \) for malware detection tasks, or a finite set for malware classification, such as \( \{1, 2, \ldots, C\} \), with \( C \) being the number of malware families.
We consider two primary data partitions: the training dataset \( \mathcal{D}_{\text{train}} \) used to train the predictive classifier, and the test dataset \( \mathcal{D}_{\text{test}} \), employed for drift assessment.
The classifier 
\( \mathbf{M} \) functionally maps the feature space to the probability space over the labels, formalized as \( \mathbf{M} : \mathcal{X} \rightarrow \mathcal{P}(\mathcal{Y}) \), and the predicted label \( \hat{y} \) for an instance \( \mathbf{x} \) is the class with the highest probability, i.e., \( \hat{y} = \argmax_{y \in \mathcal{Y}} \mathbf{M}(\mathbf{x})[y] \).

\bfnoindent{\circlednum{1} Drift Detector}
The drift detector, denoted by \( \mathbf{D} \), is tasked with quantifying the extent of drift in the test dataset \( \mathcal{D}_{\text{test}} \), with respect to the model \( \mathbf{M} \) trained on \( \mathcal{D}_{\text{train}} \). 
The detection process utilizes two main functions, i.e., uncertainty estimation function \( unc: \mathcal{X} \times \mathbf{M} \rightarrow \mathbb{R} \) and nonconformity scoring function \(ncm: \mathcal{X} \times \mathcal{Y} \times \mathcal{D}_{\text{train}} \rightarrow \mathbb{R} \).

The function \( unc(\mathbf{x}; \mathbf{M}) \) represents the uncertainty estimation associated with \( \mathbf{M} \) for an input instance \( \mathbf{x} \), outputting an uncertainty metric to reflect the confidence of the predictive model. 
Concurrently, the nonconformity scoring function \( ncm(\mathbf{x}, \hat{y}; \mathcal{D}_{\text{train}}) = dis(\mathbf{x}, sel(\hat{y}, \mathcal{D}_{\text{train}})) \), which employs a specific distance measure to evaluate how much a new observation \( (\mathbf{x}, \hat{y}) \) deviates from the calibration data selected from \( \mathcal{D}_{\text{train}} \). 
Integrating these two measures, the drift detector \( \mathbf{D} \) is defined by the following operation:
\begin{equation}
   \mathbf{D}(\mathbf{x}; \mathbf{M}, \mathcal{D}_{\text{train}}) := agg(unc(\mathbf{x}; \mathbf{M}), ncm(\mathbf{x}, \hat{y}; \mathcal{D}_{\text{train}})), 
   \label{equ:detector}
\end{equation}
where \( agg \) is a fusion function that combines the uncertainty and nonconformity scores into a singular drift metric. 
A higher output from \( \mathbf{D} \) indicates a more pronounced drift, signaling the potential necessity for model adaptation.

\bfnoindent{\circlednum{2} Drift Explainer}
The drift explainer, denoted by \( \mathbf{E} \), elucidates the features that contribute to the transition from in-distribution~(ID) data to out-of-distribution~(OOD) data. 
For a given drifting sample \( \mathbf{x}_{\text{drift}} \), the drift explainer seeks to learn a binary feature importance mask \( \mathbf{m} \in \{0, 1\}^{m \times n}\).

This involves a perturbation function, \( per: (\mathbf{x}, \mathbf{m}) \mapsto \mathbf{x}' \), that applies the mask \( \mathbf{m} \) to the sample \( \mathbf{x}_{\text{drift}} \), resulting in the perturbed sample \( \mathbf{x}_{\text{drift}}' \); a deviation function \( dev: (\mathbf{x}', \mathcal{D}) \mapsto \mathbb{R} \) that quantifies the discrepancy between \( \mathbf{x}_{\text{drift}}' \) and the training data distribution \( \mathcal{D}_{\text{train}} \).
The optimization task is defined as:
\begin{equation}
    \min_{\mathbf{m}} \left\{ dev(per(\mathbf{x}_{\text{drift}}, \mathbf{m}), \mathcal{D}_{\text{train}}) + \alpha_s reg(\mathbf{m}) \right\}.
    \label{equ:explainer}
\end{equation}
In this formulation, \( \alpha_s \) represents the regularization parameter promoting sparsity in the binary mask \( \mathbf{m} \). The regularization function \( reg \) might implement sparsity-inducing techniques such as the L1 norm or elastic-net~\cite{zou2005regularization}. 
The primary aim of this optimization is to minimize the deviation metric, ensuring that the positive values in the resulting mask \( \mathbf{m} \) pinpoint the features driving the concept drift.

\bfnoindent{\circlednum{3} Drift Adaptor}
The drift adaptor, denoted as \(\mathbf{A}\), integrates the newly annotated data into the model updating process. 
It can be conceptualized as a function 
\begin{equation}
    \mathbf{A}: (\mathbf{x}_{\text{drift}}, l_{\text{new}}, \mathcal{D}_{\text{train}}, \mathbf{M}) \mapsto \mathbf{M}'
    \label{equ:adaptor}
\end{equation}
that takes the new labels from the human annotator \( l_{\text{new}} := \mathbf{H}(\mathbf{x}_{\text{drift}}) \), the original training dataset \( \mathcal{D}_{\text{train}} \), and the current model \( \mathbf{M} \), to update the model. 

In this process, the annotated samples \( (\mathbf{x}_{\text{drift}}, l_{\text{new}}) \) are incorporated into the training dataset, resulting in an expanded dataset \( \mathcal{D}_{\text{train}}' \). 
Then, the model \( \mathbf{M} \) is retrained using this updated dataset, yielding an adapted model \( \mathbf{M}' \). 
Specifically, in the scenario of inter-class drift, the role of \( \mathbf{A} \) extends to updating the original label set \( \mathcal{Y} \), accommodating new malware classes. This necessitates a modification in the model's output layer to align with the updated label set \( \mathcal{Y}' \) before retraining. 

\section{Characterizing Current Drift Detectors} \label{sec:formalized_detectors}
Current research in malware concept drift has predominantly concentrated on the development of effective detectors. 
In this section, we examine these detectors through the lens of our proposed formalization, categorizing them based on: 
\begin{itemize}
    \item Model sensitivity: the alignment of the drift detector with the specific characteristics of the classifier, which can lead to a more precise response to model-specific drifts.
    \item Data autonomy: the detector's capability to operate independently of the training data during its operational phase, indicating adaptability and efficiency. 
\end{itemize}
For DNNs, a straightforward uncertainty measurement (\(unc\)) is probability-based, typically using the negated maximum softmax output:
\begin{equation}
    u_0(\mathbf{x};\mathbf{M}) := -\max_{y\in\mathcal{Y}}\mathbf{M}(\mathbf{x})[y].
    \label{equ:vanilla_detector}
\end{equation}
In model-sensitive drift detectors, this uncertainty measure is incorporated into the drift scoring function using various approaches. 
The design of the nonconformity score (\(ncm\)) in current detectors all involves the utilization of the classifier's training data during the testing phase. 
This includes comparing specific uncertainty values, establishing class centroids, or identifying the nearest neighbors in the latent space.

\bfnoindent{Transcendent Detector}
In Transcendent's drift detection approach, the nonconformity scoring~(\(ncm\)) is implemented using p-values~(\(dis\)) through a k-fold cross validation~\cite{vovk2015cross} approach~(\(sel\))~\footnote{The cross-conformal evaluator has the best performance within Transcendent, and we follow~{\hypersetup{citecolor=black}\cite{chen2023continuous}} to use this configuration for consistent application.}. 
For a test instance \( \mathbf{x} \), the p-value in a given fold is defined as the proportion of instances in the calibration set, which are predicted to be in the same class as \( \mathbf{x} \) and have an uncertainty score at least as high as it:  
\(
\frac{|\{\alpha \in \mathcal{C}_i[\hat{y}_i] : u_0(\alpha; \mathbf{M}_i) \geq u_0(\mathbf{x}; \mathbf{M}_i)\}|}{|\mathcal{C}_i[\hat{y}_i]|}
\).
Here, \( \mathcal{C}_i \subset \mathcal{D}_{\text{train}} \) is the calibration set for the \( i \)-th fold in the k-fold partitioning, and \( \mathbf{M}_i \) is the model retrained on the remaining training data \( \mathcal{D}_{\text{train}} \setminus \mathcal{C}_i \). 
The function \( u_0 \) measures the model uncertainty of \( \mathbf{M}_i \) for both the test instance \( \mathbf{x} \) and each calibration instance in the class \( \hat{y}_i = \mathbf{M}_i(\mathbf{x}) \).
In this case, the function \(agg\) initially combines \(unc\) implicitly in \(ncm\) within each fold, and then aggregates these results across all folds with a median-like approach.

This method is characterized as semi model-sensitive as it involves retraining models \(\mathbf{M}_i\) for each fold, instead of directly using the original classifier \(\mathbf{M}\). 
It is highly dependent on the entire training dataset, as the test uncertainties are compared with each specific training point during calibrations.

\bfnoindent{CADE Detector}
The CADE detector's innovation lies in its distance function (\(dis\)), leveraging an autoencoder to map data from \(\mathcal{X}\) into a latent space \(\mathcal{Z}\) where inter-class relationships are captured. 
In this space, the selection function (\(sel\)) identifies class centroids as the average of latent vectors for each class: \(\mathbf{c}_y = \mathbb{E}(\mathcal{Z}_{\text{train}}[y])\). 
The nonconformity measure (\(ncm\)) for a test instance \( \mathbf{x} \) is then the minimum Euclidean distance from its latent representation \( \mathbf{z} \in \mathcal{Z} \) to these class centroids \(\{\mathbf{c}_y;y\in\mathcal{Y}\}\). 
The autoencoder is trained with two loss items: the reconstruction loss to ensure the preservation of data integrity during encoding and decoding: 
\begin{equation}
   \mathcal{L}_{\text{rec}} = \mathbb{E}_{\mathbf{x}} \| \mathbf{x} - \hat{\mathbf{x}} \|_2^2 ,
   \label{equ:recon}
\end{equation}
and the contrastive loss that minimizes distances between instances of the same class and enforces a margin \( m \) between those of different classes: 
\begin{equation}
    \mathcal{L}_{\text{sep}} = \mathbb{E}_{\mathbf{x}_i,\mathbf{x}_j} \left[ \mathbb{I}_{y_i=y_j} d_{ij}^2 + \sim \mathbb{I}_{y_i=y_j} \max(0, m - d_{ij})^2 \right],
    \label{equ:sep}
\end{equation}
where \( d_{ij} = \| \mathbf{z}_i - \mathbf{z}_j \|_2 \) is the Euclidean distance between two latent representations, \( \mathbb{I}_{y_i=y_j} \) is the binary indicator that equals $1$ for same-class pairs and $0$ for different-class pairs.

The CADE detector functions independently of the classifier, with both \(unc\) and \(agg\) not explicitly defined, resulting in a lack of model sensitivity. 
This aspect may be problematic in situations where model-specific drift detection is crucial for maintaining accuracy. 
On another note, this method is semi-independent of training data, which enhances the efficiency by relying on class centroids for drift detection rather than the entire dataset. 


\bfnoindent{HCC Detector}
Adapting the HCC detector for model-agnostic applications, the learning mechanism for the distance function~(\(dis\)) closely resembles that of the CADE detector, except that: 1)~the autoencoder's decoder is replaced with a surrogate classifier \(\mathbf{M}_s\) operating in the latent space; 2)~the reconstruction loss is replaced with the binary cross-entropy loss \(\mathcal{L}_{\text{ce}}\) of the surrogate classifier and the contrastive loss is adapted to a hierarchical form \(\mathcal{L}_{\text{hc}}\), specific to intra-class drifts.
For the selection function~(\(sel\)), the HCC detector employs a nearest neighbor search in the training data. 
The final drifting score is calculated with a novel method, named pseudo loss, that uses the prediction label as the pseudo label for each test sample to calculate the losses:
\begin{equation}
    \mathbf{D}_{\text{HCC}}(\mathbf{x}; \mathcal{D}_{\text{train}}) = \hat{\mathcal{L}}_{\text{ce}}(\mathbf{x}; \mathbf{M}_s) + \beta \hat{\mathcal{L}}_{\text{hc}}(\mathbf{x}, \mathbf{x}_j;\mathbf{x}_j \in \mathcal{N}(\mathbf{x})),
    \label{equ:hcc_sampler}
\end{equation} 
where \(\mathcal{N}(\mathbf{x}) \subseteq \mathcal{D}_{\text{train}}\) is the set of nearest neighbors of \(\mathbf{x}\) in the latent space. 
The first item can be interpreted as a variation of \(u_0(\mathbf{x};\mathbf{M}_s)\) since the cross-entropy is also probability-based, and the second item is an implementation of \(ncm\) restricted by class separation. 
Therefore, the \(agg\) in this context is a weighted sum, with the weight given by the scalar \(\beta\).

In this adapted context, the HCC detector exhibits semi model sensitive as it depends on the \(unc\) derived from the surrogate classifier \(\mathbf{M}_s\).
Additionally, its operation is characterized by a reliance on the entire training dataset for identifying nearest neighbors. 


\begin{table}[tb]
    \centering
    \begin{smalltabularx}{\linewidth}{CCCCCCC}
    \toprule
           & \textbf{2016}  & \textbf{2017}  & \textbf{2018}  & \textbf{2019}  & \textbf{2020}  & \textbf{Avg.}  \\ \hline
    Prob. & 0.631 & 0.646 & 0.600 & 0.696 & \textbf{0.662} & 0.647 \\ 
    Trans. & \textbf{0.733} & 0.674 & 0.587 & 0.708 & 0.599 & 0.660 \\
    CADE & 0.721 & \textbf{0.686} & \textbf{0.739} & 0.594 & 0.487 & 0.645 \\
    HCC & 0.708 & 0.649 & 0.605 & \textbf{0.719} & 0.654 & 0.667 \\ \hline
    $\text{HCC}_{\text{ce}}$ & 0.711 & 0.649 & 0.606 & \textbf{0.719} & 0.655 & \textbf{0.668} \\
    $\text{HCC}_{\text{hc}}$ & 0.581 & 0.609 & 0.592 & 0.678 & 0.657 & 0.623 \\ \bottomrule
    \end{smalltabularx}
    \caption{Intra-class drift detection performance of existing works. For the state-of-the-art method HCC, we also make a separate study of the two items in its drift scoring function.}
    \label{tab:investigate-intra-drift}
\end{table}

\begin{table}[t]
    \centering
    \begin{smalltabularx}{\linewidth}{cCCCCCC}
    \toprule
     & \textbf{2016} & \textbf{2017} & \textbf{2018} & \textbf{2019} & \textbf{2020} & \textbf{Avg.} \\
    \hline
    NCE & 0.734 \newline \perc{3.28} & 0.702 \newline \perc{8.18} & 0.727 \newline \perc{20.04} & 0.733 \newline \perc{1.85} & 0.648 \newline \perc{-1.04} & 0.709 \newline \perc{6.11} \\ 
    \bottomrule
    \end{smalltabularx}
    \caption{The detection AUC of the proposed NCE compared with the original ce-based detector in HCC.}
    \label{tab:intra-drift-nce}
\end{table}

\section{Lessons from Current Detectors} \label{sec:motivation_lesson}
Despite the introduction of methods for drift detection in existing research, a comprehensive comparison focusing exclusively on their drift detection performance remains absent. 
This shortfall leads to a gap in understanding the true effectiveness of these methods under different conditions.
Notably, the most recent work HCC has been recognized for its effectiveness in updating malware detection classifiers.
This method is exclusively applicable to intra-class drifts.
To address the gap, our study begins by examining the problem within an intra-class scenario.
In the following, we establish a methodology for evaluating intra-class drift detection performance and provide several insights.

\bfnoindent{Evaluation Method}
For malware detection tasks, we leverage malware samples from the Malradar dataset~\cite{wang2022malradar}, which covers 4,410 malware across 180 families including singletons. 
Moreover, we make efforts to collect 43,641 benign samples in the same year period from 2015 to 2020. Our dataset has been carefully designed to mitigate spatial bias and temporal bias~\cite{pendlebury2019tesseract}. 
For the classifier, we use the \textsc{Drebin} features and an MLP model~(detailed in \autoref{sec:eva_setup}).
Regarding baselines, we take into account the vanilla probability-based detector (as depicted in~\autoref{equ:vanilla_detector}) and the three innovative detectors formalized in~\Cref{sec:formalized_detectors}.
During the evaluation phase, we assign positive drift labels to samples where predictions are incorrect, following the rationale that these misclassified samples are particularly valuable for adapting the classifier. This approach differs from inter-class drift detection evaluation, where positive drift labels are typically assigned based on the presence of unseen families, as utilized in existing work~\cite{yang2021cade}.
Then we compute the AUC metric using drifting scores and the ground truth drift labels.

\bfnoindent{Key Findings} Our evaluation results are shown in~\autoref{tab:investigate-intra-drift}, from which we have identified four key findings. 
1)~The performance order in our experiments, i.e., HCC, Transcendent, and Probability all outperforming CADE, highlights the essential role of \textit{model sensitivity} in effective drift detection. 
2)~Although CADE is not specifically tailored for intra-class drift, its superior performance in a non-model sensitive comparison with the $\text{HCC}_{hc}$ indicates the potential advantages of \textit{data autonomy} and its \textit{autoencoder-based structure}.
3)~HCC surpassing Transcendent in performance indicates that \textit{learning a distance metric}, rather than relying on calibration, can be more effective in complex models like DNNs. 
4)~The performance decline over time, particularly noticeable in the CADE method, suggests that \textit{continuous adaptation} might be necessary not just for classifiers, but also for detection models.